\documentclass[preprint,12pt]{elsarticle}

\usepackage{amsmath,amssymb,amsfonts}
\usepackage{graphicx}
\usepackage{textcomp}
\usepackage{xcolor}
\def\BibTeX{{\rm B\kern-.05em{\sc i\kern-.025em b}\kern-.08em
    T\kern-.1667em\lower.7ex\hbox{E}\kern-.125emX}}
\usepackage[normalem]{ulem}
 \usepackage{enumerate}
\usepackage{paralist}
\usepackage{tabularx}
\usepackage{booktabs}
\usepackage{comment}
\usepackage{blindtext}
\usepackage[colorinlistoftodos,prependcaption,textsize=tiny]{todonotes}
\usepackage{regexpatch}
\makeatletter
\xpatchcmd{\@todo}{\setkeys{todonotes}{#1}}{\setkeys{todonotes}{inline,#1}}{}{}
\makeatother
\usepackage{hyperref}
\usepackage[font=footnotesize,labelfont=bf]{caption}
\usepackage{array,multirow,graphicx}
\usepackage{lipsum}
\usepackage{mathtools}
\DeclarePairedDelimiter{\ceil}{\lceil}{\rceil}
\usepackage{dirtree}

\usepackage{makecell}
\usepackage{tabularx}

\usepackage[autolanguage]{numprint}
\newcommand{\uu}[2]{{\numprint{#1}}\,{$\text{{#2}}$}}

\usepackage{cleveref}
\usepackage{dblfloatfix}
\usepackage{nicefrac}
\usepackage[caption=false]{subfig}
\usepackage[draft,inline,nomargin,index]{fixme}
\usepackage{mdframed}
\usepackage{tikz}
\usepackage{pifont}
\usepackage{float}
\usepackage{amsmath,amsfonts}
\usepackage{amssymb}
\graphicspath{ {./images/} }

 \definecolor{monrouge}{HTML}{AD4B42}
 \definecolor{red}{HTML}{FF0000}
 \definecolor{monbleu}{HTML}{4F6A9C}
 \definecolor{monvert}{HTML}{2da44e}
 \definecolor{monjaune}{HTML}{DCB355}
 \definecolor{moranbl}{HTML}{90A5A7}
 \definecolor{cyan}{HTML}{00AAAA}
 \newcommand{\sebastian}[1]{{\color{monrouge}{\bf SEBASTIAN:} \small #1}}

\newcommand{\blindcomment}[1]{}
\newcommand{\hide}[1]{}
\newcommand{\yshide}[1]{}

\usepackage{tcolorbox}
\usepackage{varwidth}
\definecolor{ForestGreen}{RGB}{34,139,34}

\newtcolorbox[list inside=mynote,auto counter,
crefname={box}{boxes}]{MyNote}[1][]{fonttitle=\fontsize{9.0pt}{8.0pt}\selectfont\bfseries\sffamily,arc=0.5mm,leftrule=0mm,rightrule=0mm,colback=ForestGreen!5!white,colframe=ForestGreen!85!black,
bottomrule=1mm, fontupper=\fontsize{11.0pt}{8.0pt}\selectfont\sffamily, title={Box \thetcbcounter}, #1}

\creflabelformat{equation}{#2#1#3}

\newtagform{noparen}{}{}
\usetagform{noparen}

\usepackage{listings}
\definecolor{ntaliceblue}{rgb}{0.94, 0.97, 1.0}
\definecolor{bleudefrance}{rgb}{1.0, 0.44, 0.37}
\definecolor{backcolour}{rgb}{0.8,0.8,0.8}
\definecolor{beige}{rgb}{0.96, 0.96, 0.86}
\usepackage{color}
\definecolor{mygreen}{rgb}{0,0.6,0}
\definecolor{mygray}{rgb}{0.5,0.5,0.5}
\definecolor{mymauve}{rgb}{0.58,0,0.82}
\definecolor{anti-flashwhite}{rgb}{0.95, 0.95, 0.96}

\usepackage{fixme}
\fxsetup{theme=color,mode=multiuser}

\FXRegisterAuthor{kk}{akk}{\color{monbleu}\textbf{KK}}
\FXRegisterAuthor{sf}{asf}{\color{monrouge}\textbf{SF}}
\FXRegisterAuthor{br}{abr}{\color{monvert}\textbf{BR}}
\FXRegisterAuthor{lb}{alb}{\color{monjaune}\textbf{LB}}
\FXRegisterAuthor{yl}{ayl}{\color{moranbl}\textbf{YL}}

\usepackage{acro}

\DeclareAcronym{pf}{
  short=PF,
  long=Particle Filtering,
}

\DeclareAcronym{enkf}{
  short=EnKF,
  long=Ensemble Kalman Filtering,
}

\DeclareAcronym{pdf}{
  short=PDF,
  long=Probability Density Function,
}

\DeclareAcronym{pfs}{
  short=PFS,
  long=Parallel File System,
}

\DeclareAcronym{nwp}{
  short=NWP,
  long=Numerical Weather Prediction,
}

\DeclareAcronym{wrf}{
  short=WRF,
  long=Weather Research and Forecasting Model,
}

\journal{Computational Science}

\begin{document}

\begin{frontmatter}

\title{A Framework for Large Scale Particle Filters Validated with Data Assimilation for Weather Simulation}




\author[1]{Sebastian Friedemann}
\ead{sebastian.friedemann@inria.fr}

\author[2]{Kai Keller}
\ead{kai.keller@bsc.es}

\author[3]{Yen-Sen Lu}
\ead{ye.lu@fz-juelich.de}

\author[1]{Bruno Raffin\corref{cor1}}
\ead{bruno.raffin@inria.fr}

\author[2]{Leonardo Bautista-Gomez}
\ead{leonardo.bautista@bsc.es}

\cortext[cor1]{Corresponding author}

 \affiliation[1]{organization={Univ. Grenoble Alpes, Inria, CNRS, Grenoble INP, LIG}, 
                 postcode={38000}, 
                  city={Grenoble}, 
                  country={France}}

  \affiliation[2]{organization={Barcelona Supercomputing Center},
                  city={Barcelona},
                  postcode={08034},
                  country={Spain}}

 \affiliation[3]{organization={Forschungzentrum Juelich},
                  city={Juelich},
                  postcode={52428},
                  country={Germany}}



\begin{abstract}

  Particle filters are a group of algorithms  to solve  inverse problems through statistical  Bayesian methods
  when the model does not comply with the linear and  Gaussian hypothesis. Particle filters
  are used in domains like  data  assimilation, probabilistic programming,  neural network
optimization, localization and navigation. Particle filters estimate the probability
distribution of model states by running a large number of model instances, the so called  particles. 
The ability to handle a very large number of particles is critical for high dimensional models.
This paper proposes  a novel paradigm to run  very large  ensembles of parallel model instances on supercomputers.  
The approach combines an elastic and fault tolerant runner/server model minimizing data movements
while enabling dynamic load balancing. Particle weights  are computed locally on each runner  and
transmitted when available to a server that normalizes them, resamples new particles  based on their weight, and redistributes dynamically the work to
runners to react to load imbalance. Our approach relies on a an asynchronously managed
distributed particle cache permitting particles to move from one runner to another in
the background while particle propagation goes on. This also enables the number of
runners to vary  during the execution either in reaction to failures and restarts, or
to adapt to changing resource availability  dictated by external decision processes.
The approach is experimented with the Weather Research and Forecasting (WRF) model, to
assess its  performance for probabilistic weather forecasting. Up to \numprint{2555}
particles on \numprint{20442} compute cores are used to assimilate cloud cover
 observations into short--range weather forecasts over Europe.
\end{abstract}




\end{frontmatter}

 \section{introduction}

Given an output and a transformation function, finding the input states
represents a so called \textbf{inverse problem}. A wide range of
approaches to address this central problem exist.
Statistical Bayesian methods stand out as they provide uncertainty
measures of the proposed input in form of probability density functions.
In this paper, we consider \textbf{particle filters}, a statistical
Bayesian method combining uncertainties of both the dynamical system and
observations, to estimate the system  \textbf{state}.  Several
realizations of the dynamical system, called \textbf{particles}, with
differently perturbed internal states, are \textbf{propagated} up to a
time where \textbf{observation data} are available. These particles are
then \textbf{weighted} corresponding to their distance to the observations.
The weights are used to generate a new sample of particles that better
matches the observations.  This process repeats while observations are
available.

Particle filters are used for several purposes, like  {\it Data  Assimilation
  (DA)}~\cite{Leeuwen-particlefiltersurvey-2019},  probabilistic programming~\cite{lunden2021correctness,ronquist2021universal,Meent-probprog:2018}, neural network
optimization~\cite{freitas-SequentialMonteCarlo-2000}, localization and
navigation\cite{blok_robot_2019}. Particle filters stands  by their ability to work with  nonlinear and/or non-Gaussian state space models in opposition to technics like \ac{enkf}. But this ability  comes with a need to run larger number of particles.  If the dynamical system is an advanced parallel high-dimensional numerical model solver, as for
geoscience applications, thousands of particles may be necessary to avoid undersampling and degeneracy. While
high-dimensional large-scale solvers are compute intense already, the execution of
several thousands of instances  adds orders of magnitude of calculations.
Large scale DA with particle filters  is for instance used  for geoscience applications such as
weather forecasting~\cite{Bauer2021}.   Supercomputers, reaching today Exascale, have the compute
power to support very large scale particle filters. But using such resources efficiently, time and energy wise, is challenging.  Applications
need to limit the use of the Parallel File System (PFS),  a classical supercomputer bottleneck,
and favor instead  in  situ data processing  as well as   local data storage  to  reduce data movements, asynchronism to
overlap tasks whenever possible.  Applications also need to be flexible  to adapt to changes  during
execution, requiring support  for resilience, elasticity and dynamics load balancing.
 
 
Existing large scale approaches can be divided into two types:
\textbf{online} and \textbf{offline} approaches.  Offline approaches use temporary files to exchange
data. To propagate one  particle, one  model instance   starts, loads  the particle
from a  file, propagates it up to a given time, stores the resulting particle back to a file and shuts
down.  This approach is flexible, fault tolerance is easy to support, but performance, especially at
scale is  impaired by the heavy use of the file system and the cost of starting a new model instance
for each propagation. Online approaches bypass the file system by running  a large MPI
application that encompasses  the full workflow, where the particles are distributed
to the different model instances through the network via MPI communications.
While saving I/O overheads, this approach  loses  flexibility. For instance, a fault during a single
particle propagation stops the entire application.  Thus existing online approaches, as will be
detailled in the related work section (\Cref{sec:related-work}), usually do not support fault
tolerance or  dynamic load balancing.
 
In  this paper  we  develop an  alternate  approach that  leads  to a  high  efficient yet  flexible
framework.  The key to achieve this goal is the \textbf{virtualization} of particle propagations. We
turn  a numerical  model  solver instance  into  a \textbf{runner}  capable  of propagating  several
particles one  after the  other with low  overheads and idle  times. Each  runner is coupled  with a
node-local distributed  state cache  enabling fast  loads and  stores of  particles. The  caches are
asynchronously persisted  to the file system  for checkpointing and load  balancing between runners.
Asynchronous  prefetching of  particles  into  the cache  enables  overlapping  particle loads  with
the particle propagation.  A \textbf{server} organizes the work distribution to the runners and performs
the centralized tasks of  the particle filter update and (re-)sampling. Runners  and server are each
executed  as   independent executables  to  support  elasticity  and  facilitate
fault tolerance.  The  association of these different  features complemented with a  fault tolerance
protocol, leads  to an  elastic and  resilient framework, minimizing  data movements  while enabling
dynamic load  balancing.  Particle virtualization enables  to decouple resource allocation  from the
number of  particles. The  number of runners  can vary  during the execution  either in  reaction to
failures and restarts, or  to adapt to changing resource availability  dictated by external decision
processes.  The proposed  architecture is  designed for  running at  extreme scale,  leveraging deep
storage hierarchies and heterogeneous cluster designs of current and future supercomputers.

We strain our proposed particle filter
framework with a realistic use-case, interfacing with the Weather
Research and Forecasting (WRF, version 3.7.1) model~\cite{Skamarock2008}. WRF is a widely used weather
model for operational forecasting and research.  Using our
particle filter, we are able to run \numprint{2555}~particles on
\numprint{20442}~compute cores for WRF simulations on a European domain
with 87~\% efficiency.
 
The rest of the paper is structured as follows: \Cref{sec:data-assimilation} reviews
the principles of particle filters and the associated workflow.
\Cref{sec:impl} presents the architecture of our proposed approach,
while \Cref{sec:exp} is dedicated to experiments and   \Cref{sec:discussion}
to discussion. The papers ends with  related work in \Cref{sec:related-work} and a conclusion
 in~\Cref{sec:conclusion}.
 
 \section{Particle Filters} \label{sec:data-assimilation}


\begin{figure}
    \centering
    \includegraphics[width=8.3cm]{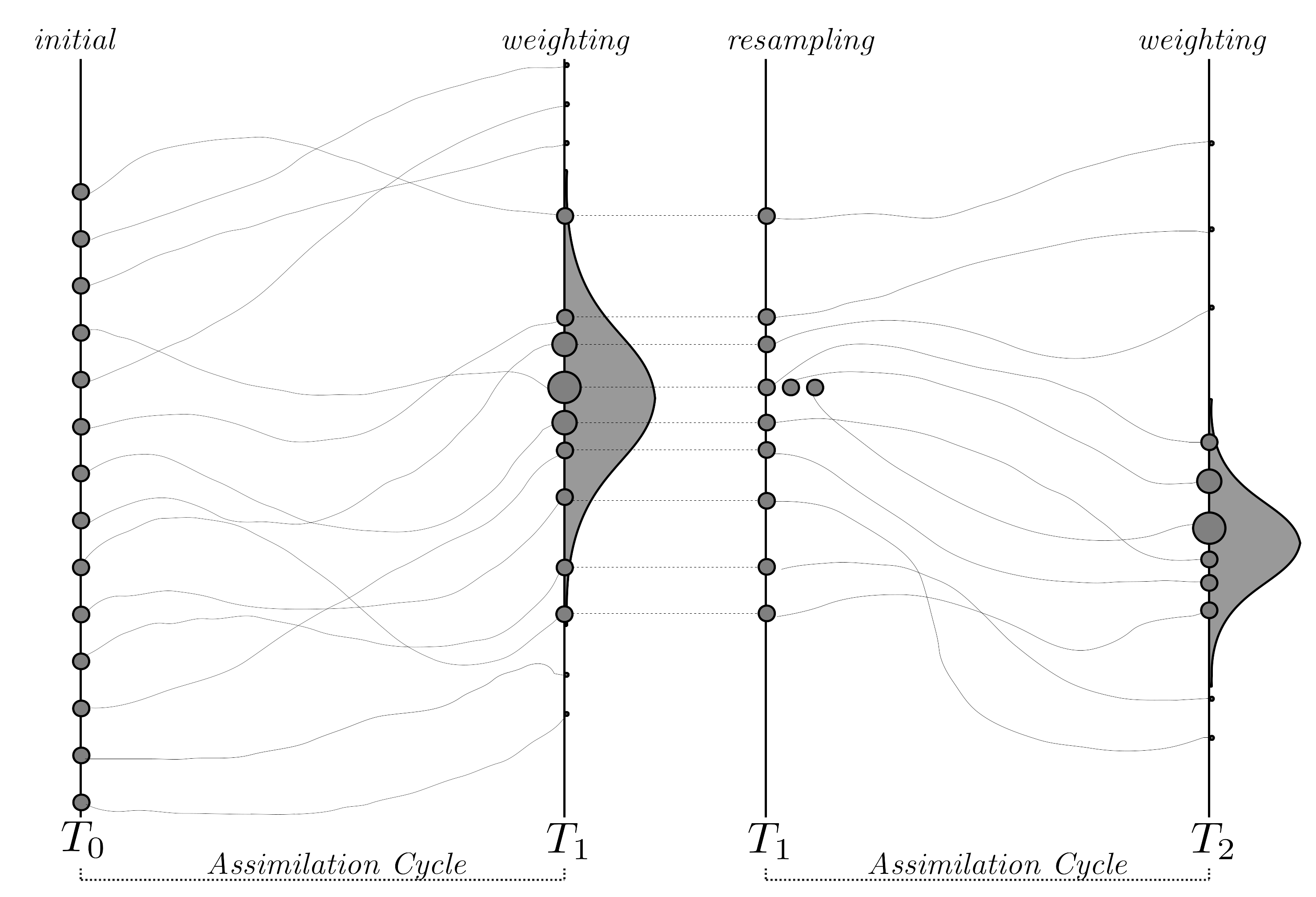}
    \caption{Initially particles are uniformly   sampled. They are propagated to $T_1$ where they
      are  weighted taking into account  observation data. Resampling leads to discard  some particles with low
      weights (top and bottom), while  others  with high weights become parent of  several
      ones (3 here).}
    \label{fig:SIR_DIAGRAM}
  \end{figure}

In this section, we give a brief introduction on the particle filter
formalism, focusing on properties that we exploit in our proposal. For a
comprehensive introduction, we refer
to~\cite{Leeuwen-particlefiltersurvey-2019,candy_bootstrap_2007}.  Let $\mathcal{M}$ be a
numerical model, that propagates a particle $p$ from state
$\textbf{x}_{p,t-1}$ at time $t-1$ to state $\textbf{x}_{p,t}$ at time
$t$:
\begin{equation}   \label{eq:model-operator}
  \textbf{x}_{p,t} = \mathcal{M}(\textbf{x}_{p,t-1})+ \boldsymbol\beta^{t}
\end{equation}
Where $\boldsymbol\beta$ is a random forcing representing errors in the model.
Let $\mathcal{H}$ be the projection operator from the state space to
the observation space:
\begin{equation}
  \textbf{y} = \mathcal{H}(\textbf{x}) + \boldsymbol\epsilon^t
\end{equation}
Where $\boldsymbol\epsilon$ is a random vector, representing the measurement
errors.

The bootstrap particle filter formalism can be derived using Bayes'
theorem:
\begin{equation}
  p(\textbf{x}_t|\textbf{y}_t) =
  \frac{p(\textbf{y}_t|\textbf{x}_t)\,p(\textbf{x}_t)}{p(\textbf{y}_t)}
  \label{eq:bayes_theorem}
\end{equation}
Where $p(\textbf{x}_t|\textbf{y}_t)$ is the posterior \ac{pdf},
$p(\textbf{x}_t)$ is the prior \ac{pdf}, $p(\textbf{y}_t|\textbf{x}_t)$ is the likelihood of observing
$\textbf{y}_t$ if $\textbf{x}_t$ would represent the true state, and
$p(\textbf{y}_t)$ is the evidence available. The goal of the filtering formalism is to
derive the posterior $p(\textbf{x}_t|\textbf{y}_t)$, which describes the \ac{pdf} of the state
$\textbf{x}_t$ taking into account the evidence $\textbf{y}_t$.

In the bootstrap particle filter, the prior $p(\textbf{x}_t)$ is
estimated via sampling an ensemble of $P$
particles $\textbf{x}_{p,t}$ representing different model states
\begin{equation}
  p(\textbf{x}_t) =  \frac{1}{P}  \, \sum_{p=0}^{P-1}\, \delta(\textbf{x}_t-\textbf{x}_{p,t}), \label{eq:bootstrap-pdf}
\end{equation}
The likelihood $p(\textbf{y}_t|\textbf{x}_t)$ is assumed to be known,
estimated when calibrating the sensor. It is 
derived from the \ac{pdf} of $\boldsymbol\epsilon$ applied
to the distance between state and observation
$\textbf{y}_t-\mathcal{H}(\textbf{x}_t)$:
\begin{equation}
  p(\textbf{y}_t|\textbf{x}_t) =
  p_{\epsilon}(\textbf{y}_t-\mathcal{H}(\textbf{x}_t))
\end{equation}
The evidence $p(\textbf{y}_t)$ can be computed by:
\begin{align}
  p(\textbf{y}_t) &= \int\, p(\textbf{y}_t|\textbf{x}_{t})p(\textbf{x}_{t})\, d\textbf{x}_{t} \\
  &= \sum_{p=0}^{P-1}\, \frac{1}{P}\,p(\textbf{y}_t|\textbf{x}_{p,t}) \\
\end{align}
Putting all together and replacing the expressions in Bayes' theorem
(\Cref{eq:bayes_theorem}) we arrive to the expression for the posterior~\cite{Leeuwen-particlefiltersurvey-2019}:
\begin{equation}
  p(\textbf{x}_t|\textbf{y}_{t}) \approx \sum_{p=0}^{P-1}\, \hat{w}_{p,t}\,\delta(\textbf{x}_t-\textbf{x}_{p,t}) \label{eq:pdf-posterior}
\end{equation}
With $\hat{w}_{p,t}$ being the \emph{normalized} particle weights:
\begin{equation}
  \hat{w}_{p,t} =
  \frac{p(\textbf{y}_t|\textbf{x}_{p,t})}{\sum_{q=0}^{P-1}\,p(\textbf{y}_t|\textbf{x}_{q,t})}
  = \frac{w_{p,t}}{\sum_{q=0}^{P-1}\,w_{q,t}} \label{eq:weight-normalized}
\end{equation}
and $w_{p,t}$ being the \emph{unnormalized} particle weights:
\begin{equation}
  w_{p,t} = p(\textbf{y}_t|\textbf{x}_{p,t})\, w_{p,t-1} \label{eq:weight-unnormalized}
\end{equation}
Note that the initial weights are set equal to $w_{p,0} = 1/P$.

Especially for high dimensional models, particle filters tend to suffer
from weight degeneration, i.e., one normalized weight is close to one and
all the others are close to zero. A classical approach against ensemble
degeneration is Sequential Importance Resampling
(SIR)~\cite{gordon_novel_1993,liu_theoretical_2001}. The procedure of
SIR consists in resampling particles from the posterior
(\Cref{eq:pdf-posterior}) at the end of the propagation step; $P$
particles are randomly drawn, \textit{resampled}, from the
existing particles, each with a probability $w_{p,t}$. Low weighted
particles become discarded, while high weighted particles can become the
starting point of multiple particle propagations
(\Cref{fig:SIR_DIAGRAM}). More precisely, the resampling leads to the multiset
$P$ defined by the ordered pair $(Q,\boldsymbol{\alpha})$. Where $Q$ is the
set of unique particles $q$ in P, and $\alpha_q$ the number of
the occurrences of $q$ in $P$. The particles $q$ are hereinafter called
\textbf{parent particles}.

The resampled particles are all assigned  the same  weight of $w_{p,t} = 1/P$ again.
Particles departing from the same parent may need to become stochastically perturbed if the model does not
contain a stochastic component itself.  Otherwise, the trajectories of
those particles  would be identical.

Different flavors of SIR and
resampling algorithms, like Residual Resampling,
exist~\cite{bolic_new_2003}. Some perform a resampling step after each
propagation phase, while others make this dependent on criteria like the
variance of the weights.  In this paper we rely on  SIR with resampling
after each propagation phase.

\begin{center}
\begin{MyNote}[label=box:properties, width=0.95\linewidth, box
  align=center, halign=center, valign=center]
  \begin{enumerate}[(a)]
    \item The propagation of particle $p$ depends only on the associated state $\textbf{x}_{p,t}$
      and can be performed independently of other particles.
    \item Weights $w_{p,t}$ depend only on the associated particle $p$ and
      observation vector $y_t$, and can be computed independently of other weights and particles.
    \item The filter update only depends on the weights $w_{p,t}$, and not on the particles and associated states. 
    \item The states $\textbf{x}_{p,t}$ associated to the particles $p$ remain unchanged during the filter update.
  \end{enumerate}
\end{MyNote}
\end{center}

\Cref{box:properties} lists the properties of particle filters that
are the basis for our implementation.

We exploit property (d): In contrast to other DA
techniques, such as \ac{enkf}, particle states
remain unchanged during the filter update. Particles that have departed
too much from the observations (low weights) are discarded, and the
sample set is narrowed around the best particles (high weights). The
associated states, however, are not changed.  Property (a) follows directly from~\Cref{eq:weight-unnormalized}.
Property (b) results from decoupling the weight calculation from the
filter update (decentralization). The update itself, only consist of
the weight normalization and particle resampling.  Finally, property (c)
is an intrinsic property of the bootstrap particle filter, since particles are either
withdrawn or selected, but not changed. In the following sections, we
will show how we can exploit those properties to
improve efficiency of and resilience particle filter implementations.



\hide{

We   introduce quickly   the   background   on  particle   filters   for   data   assimilation.    Refer
to~\cite{Leeuwen-particlefiltersurvey-2019} for an extensive survey  on the topic.

Particle filters
use a set of  particles to  represent the probability distribution
of a process state given noisy and partial input or models relying on heuristics. Particle filters distinguish themselves from other DA
methods like EnKF~\cite{Evensen-DABook-2009,evensen_sequential_1994,Houtekamer-DA-ENKF-1998} or 3/4DVar~\cite{lorenc_met_2000,kooij-connally_technical_nodate} by supporting flow dependent errors, nonlinear state/space models, and the distributions of observation errors and
initial  states can  follow  any  probability distribution.  These properties  motivate   their use
for weather forecasting~\cite{berndt_predictability_2018} similar to the experiments
in this paper.

Let $\mathcal{M}$ be a numerical model, that \textit{propagates} a particle, i.e., its according particle state $\textbf{x}$ in time. It computes the state at the next time step~$x_t$ from a particle state~$x_{t-1}$:
\begin{equation}
   x_{t} = \mathcal{M}(x_{t-1}) 
 \end{equation}

Let  $y_t$ be an observation obtained by measures on the real system (satellite images, buoys, ground measuring stations, etc.).  This data is often not homogeneous to the particle state~$\textbf{x}$ that is handled by the numerical model. The
 operator~$\mathcal{H}$ projects a particle state $\textbf{x}$ into observation space.  We consider that
 the observation is the projection of the true state plus an error~$\epsilon$ whose distribution~$p_\epsilon$ is known:

\begin{equation}
    y_t = \mathcal{H}(\textbf{x}^\text{true}_t) + \epsilon_t
\end{equation}

DA acts in \textit{assimilation cycles} (\Cref{fig:SIR_DIAGRAM}). First the $M$ particles $x_{i,t-1}$ are propagated through the model~$\mathcal{M}$
up to the next time step when observations are available ($t$ for simplicity here). Then the normalized particle weight $\hat{w}_{i,t}$ that approximates the
probability  $p(x_{i,t} | y_{t})$ is computed for each particle state~$x_{i,t}$:

\begin{equation}
    \label{eq:normalizedweight}
\hat{w}_{i,t} = \frac{1}{M} \frac{p(y_{t} | x_{i,t})}{p(y_t)}  \approx  \frac{w_{i,t}}{\sum_{0 \leq j < M} w_{j,t}},
\end{equation}
where
\begin{equation}
 w_{i,t} = p(y_{t} | x_{i,t})=p_{\epsilon_{t}}( y_{t}-\mathcal{H}(x_{i,t})).
\end{equation}
Finally, the probability density function of $p(x_{t} | y_{t})$ is approximated by:
\begin{equation}
  p(x_{t} | y_{t})=\sum_{0 \leq i <M}  \hat{w}_{i,t} \delta(x_{t}-x_{i,t}),
\end{equation}
where $\delta$ is the \textsc{Dirac} measure.

Especially for high dimensional problems, particle filters tend to suffer from weight degeneration, i.e., one normalized weight is
close to one and all the others to zero,  making the particle sample meaningless. To avoid this issue,
one  classical  approach  consists in  resampling  the  particles based  on their importance (SIR,
Sequential Importance Resampling) before starting the next cycle.  A new ensemble of $M$ particles
is drawn, each one with a probability $\hat{w}_{i,t}$. This leads to discard low weight particles
while high  weight ones can become  the {\it parent} of  multiple new particles
(\Cref{fig:SIR_DIAGRAM}). Particles from the same parent particle state  may need to be slightly perturbed
if the model does not  contain a stochastic component, so they do not propagate to the same state.


}

 \section{Architecture}\label{sec:impl}


\begin{figure}
    \centering
    \includegraphics[width=0.9\linewidth]{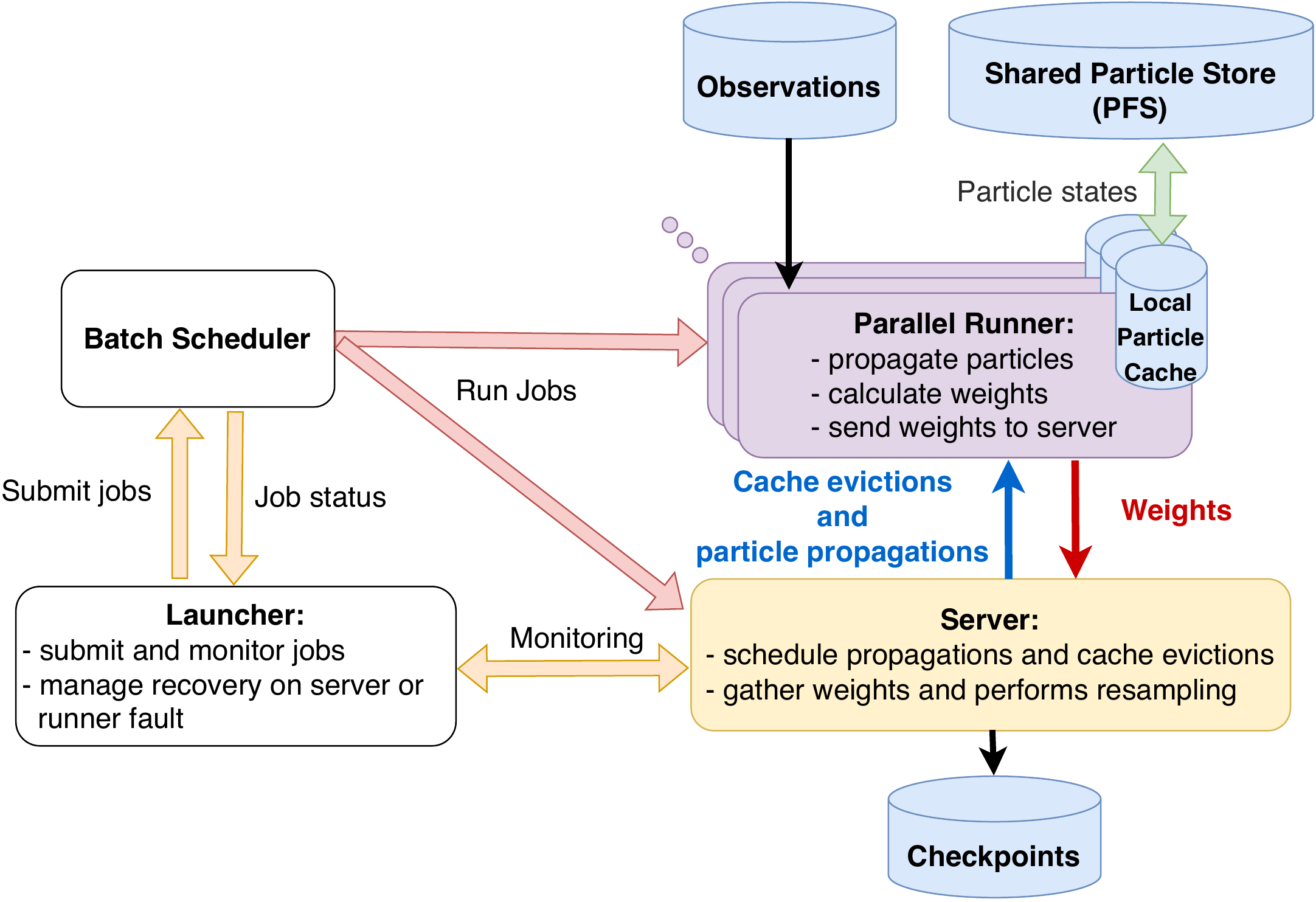}
    \caption{Architecture overview.}
    \label{fig:architecture-overview}
  \end{figure}


In this section we detail the proposed architecture to run  a large number of particles
with  parallel numerical models. The algorithm, as  presented in \Cref{sec:data-assimilation},
is a sequence of two main steps:
\begin{enumerate}
\item A first compute intensive massively parallel step where particles can be processed concurrently to:
    \begin{enumerate}
  \item \label{item:part-prop} propagate each sate:  $\textbf{x}_{p,t}=\mathcal{M}(\textbf{x}_{p,t-1})$,
  \item  \label{item:weight-comp} compute each  unormalized weight  from each state and   observation data:
    \begin{align}
      w_{p,t}   &=  p_{\epsilon}(\textbf{y}_{t}-\mathcal{H}(\textbf{x}_{p,t}))
    \end{align}
  \end{enumerate}
\item A second  lightweight step  that requires to gather all unormalized weights   $w_{p,t}$, usually one double per weight, for normalization and resampling.
\end{enumerate}
We attribute the first step work to  runners and the second step to a  \textbf{server}.
A \textbf{runner} is designed  to propagate  several particles one  after the  other with low  overheads and idle  times (\Cref{fig:architecture-overview}).  Each one is coupled  with a node-local  distributed cache  enabling fast  loads and  stores of  particles. The  caches are asynchronously persisted  to the global file system  for checkpointing and dynamic load balancing (i.e., ensure global availability of the particles).  Because resampling can lead to discard some particles, or duplicate others  originating  from the same parent (with a local perturbation if needed), states need to be dynamically redistributed to runners to keep them evenly busy.  The server drives the dynamic  distribution of particle propagation  tasks to runners. Runners use the distributed cache to load from the file system the missing states.   This design ensures low communications between the server and runners, and reduced state movements.  The runners  and  the server run   as   independent executables,  enabling to have a dynamically changing  number of runners. This is a key feature used for  fault tolerance and elasticity.  Elasticity (sometimes also called maleability) is the  ability to run  under changing resource availability, here varying number of runners.

In the following we detail this design: the runners
(\Cref{sec:runners}), the server (\Cref{sec:server}), the distributed cache (\Cref{sec:distributed-cache}), the workflow between these components (\Cref{sec:workflow}), the particle propagation scheduling (\Cref{sec:implementation:scheduling}), the jobs monitoring (\Cref{sec:implementation:launcher}),
and the fault tolerance protocol (\Cref{sec:fautltolerance}) before ending with additional implementation details (\Cref{sec:implementation:details}).

\subsection{Runners}\label{sec:runners}

Runners are built from the  simulation code,  often an advanced parallel
code  or even a coupling of several parallel codes, with significant
start-up times to load and build the different internal data structures.
To avoid  paying the cost of a restart for each particle propagation, we
augment the simulation code with a mechanism to store and load particle
states. This is the base of  \textbf{particle virtualization}:
a  runner can load a  particle, propagate it, store the
result, and repeat this as many times as necessary.
Runners are  associated with a distributed cache to accelerate state
loads and stores as detailled in \Cref{sec:distributed-cache}.
Runners also compute the associated weights $w_{p,t}$. Hence, each  runner also
needs to load the observations $\textbf{y}_t$ once per cycle.  Notice
that the size of the observations is typically much smaller than the
size of the states $\textbf{x}_{p,t}$.



\subsection{Server}\label{sec:server}

The server is entrusted with multiple tasks. First, it is responsible
for scheduling the particle propagations to the runners (\Cref{sec:implementation:scheduling}). Second, it
gathers the weights from the runners and performs the resampling at the
end of each assimilation cycle. Third, it controls the content of a
distributed particle cache (\Cref{sec:distributed-cache}). To collect the weights $w_{p,t}$, the
server is messaged from the runners after each propagation. If there are
still particles to propagate in the current cycle, the server responds to the message with an
id uniquely defining a particle (hereinafter called particle-id) for the next propagation. If not, the server
performs the resampling and starts the new cycle by scheduling the sampled
particles to the runners. Very little data is exchanged between a runner and server.
The runners send the particle-id (a single int) and the corresponding weight (a single float), and the server responds
with the particle-id next to propagate.

\subsection{Distributed Particle Cache}\label{sec:distributed-cache}

\begin{figure}
    \centering
    \includegraphics[width=1.0\linewidth]{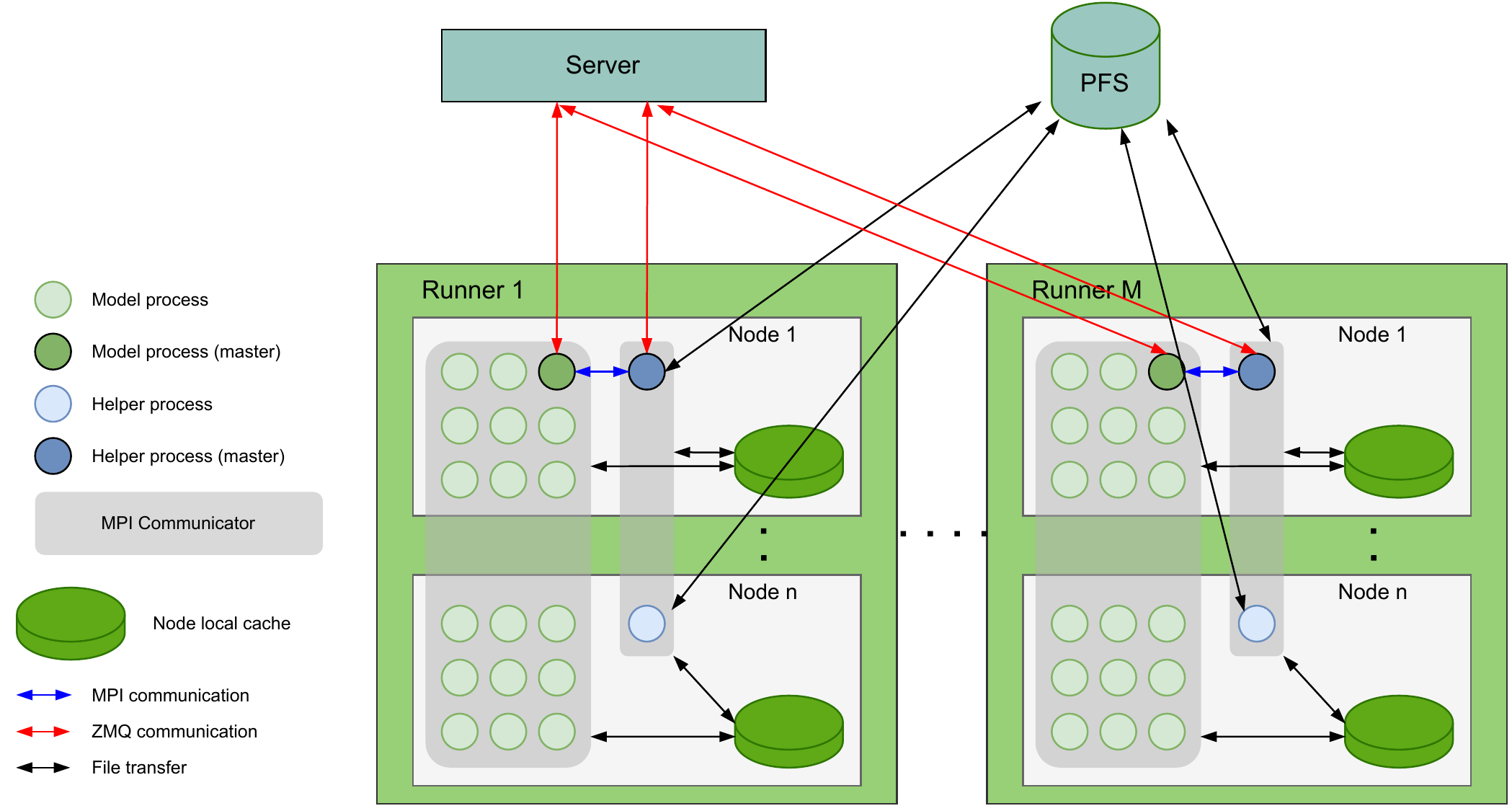}
    \caption{Internal runner  architecture and interactions with the server and global storage (PFS).
      Communications  with the server combine MPI and ZMQ data exchanges.}
    \label{fig:components:runner-server}
\end{figure}

\begin{figure}
    \centering
    \includegraphics[width=1.0\linewidth]{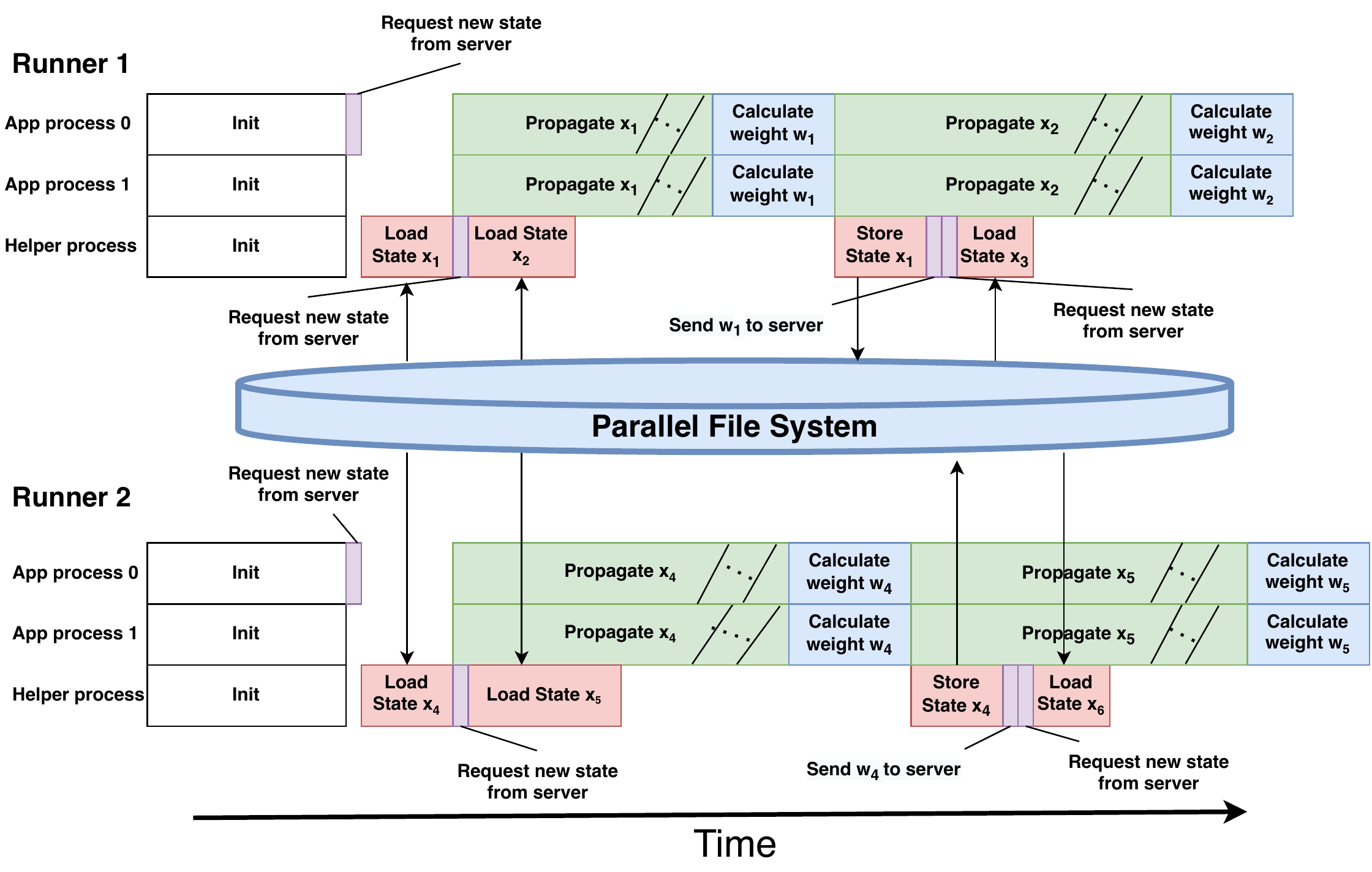}
    \caption{Schematic Gantt diagram showing the activity of  two runners (initialization followed
      by  two    assimilation cycles).  Focus on how the  helper process asynchronous loads and
      stores enables to shadow the parallel file system accesses. For sake of simplicity no
      cache is used here.}
    \label{fig:runners-gantt}
\end{figure}


To allow multiple propagations of one  particle on different runners, it is necessary to make them globally available.  A straight  forward approach  is to store  particles on global  storage. However,  on supercomputers
global storage is subject  to large throughput variability due to the high  workload and the limited
bandwidth. Node-local storage, on the  other hand, is  only used by  the processes that run  on the
nodes,  and the  bandwidth  can be  stacked.  Storing the  particles locally  results  in scalable  I/O
performance, scaling linearly with the number of nodes.

To leverage node-local storage while still providing the particles globally, runners rely on  a
\textbf{distributed particle cache}. Each runner executes \textbf{helper processes} (one per node)  in addition to the  \textbf{model
 processes}, where both groups of processes are associated with its own MPI communicator
(\Cref{fig:components:runner-server}). The model processes propagate the particles and store the
associated states locally on the nodes allocated to the runner (RAM~disk or other node-local storage when
available). The helper processes  then stage the states from local to global storage asynchronously,
enabling to overlap the associated I/O costs (\Cref{fig:runners-gantt}). Also notice that persisting particles to  global storage acts  as a particle checkpoint used
by the   fault tolerance protocol (\Cref{sec:fautltolerance}).

We allow keeping a number of particles in each of the runner caches  to exploit property (d)
from~\Cref{box:properties}: resampling does not change the particle states. Hence, keeping propagated particles in the cache, increases the probability to find a particle locally for future propagations (i.e., during the next cycle). If available in its local cache,  a runner can propagate a particle without loading it from global
storage.  To further minimize cache loads,  runners implement an optimized cache eviction strategy. The eviction strategy becomes especially important if the cache capacity is exceeded by the accumulated size of the particles propagated during one cycle. Because the runners have no knowledge about the status of the particle filtering (propagations, resampling), the evictions are controlled by the server and directed to the runners.

As  explained in~\Cref{sec:workflow}, each time a particle  has been stored
in the cache by the model processes upon  successful propagation, the
helper processes copy it in the background to global storage. Hence, all propagated particle states can be selected for eviction, since they are safely stored on global storage. When an eviction is required,
the server selects a particle from the cache in the following order:
\begin{enumerate}
  \item A  particle from the previous cycle  discarded by resampling;
  \item A parent particle from the current cycle for which all associated propagations have been performed, and all weights received;
  \item The particle with the lowest weight propagated during the current cycle;
  \item A  randomly selected particle.
\end{enumerate}
The particle states for cases 1 and 2 can safely be removed from  cache, since
those particle are  not  needed anymore for future propagations. In case 3, we select the particle state with the lowest weight, as it has the lowest probability of being selected for future cycles during the resampling.

\subsection{Runners/Server Workflow}\label{sec:workflow}

Once a runner job has started, it dynamically connects to the server and  requests a particle
to propagate from it. The server selects the particle following a scheduling policy described  in~\Cref{sec:implementation:scheduling}.
The  model checks the location of the particle. If already located inside the local cache, the
propagation starts.  Otherwise, the  model processes request the
helper processes to load the state into the cache.
The model processes block until the helper processes fetched the
particle into the cache, and afterwards start the propagation.

Once a  particle propagation finishes, the model computes the associated
weight $w_p$ and stores the particle into the cache. Further, the weight and particle-id are sent to the helper processes and a new particle is requested for propagation. The helper processes, after
receiving the weight from the model processes, stage the particle from
the cache to global storage and afterwards sends the weight and particle-id to the server. This order ensures that  the server receives a weight only after the corresponding particle is propagated and successfully stored on global storage.

The helper processes further prefetch particles in parallel to the propagations (\Cref{fig:runners-gantt}).
The goal is to avoid blocking the model processes while waiting for a particle load from global storage (cache miss). Each time  helper processes send  a weight to the server, they also request the next-to-next particle-id to propagate. This particle is prefetched into the cache to become
locally available for the next to follow propagation. Prefetching is suspended at
the end of each  propagation cycle, as propagation work for the next cycle becomes only known after the server has performed the resampling of all particles. Notice that a   helper may need to cancel prefetching
if the prefetched particle was in the meantime assigned to another runner, making idle the model process while waiting for the next particle to propagate.
When the server  makes  such  a decision to  better balance the work load, it also  takes care of ensuring coherency
between runners.  Globally prefetching proved to be very efficient for overlapping particle state loads with propagation (\Cref{sec:overlapexp}).

\subsection{Particle Propagation Scheduling}\label{sec:implementation:scheduling}

\begin{figure}
    \centering
    \includegraphics[width=1.0\linewidth]{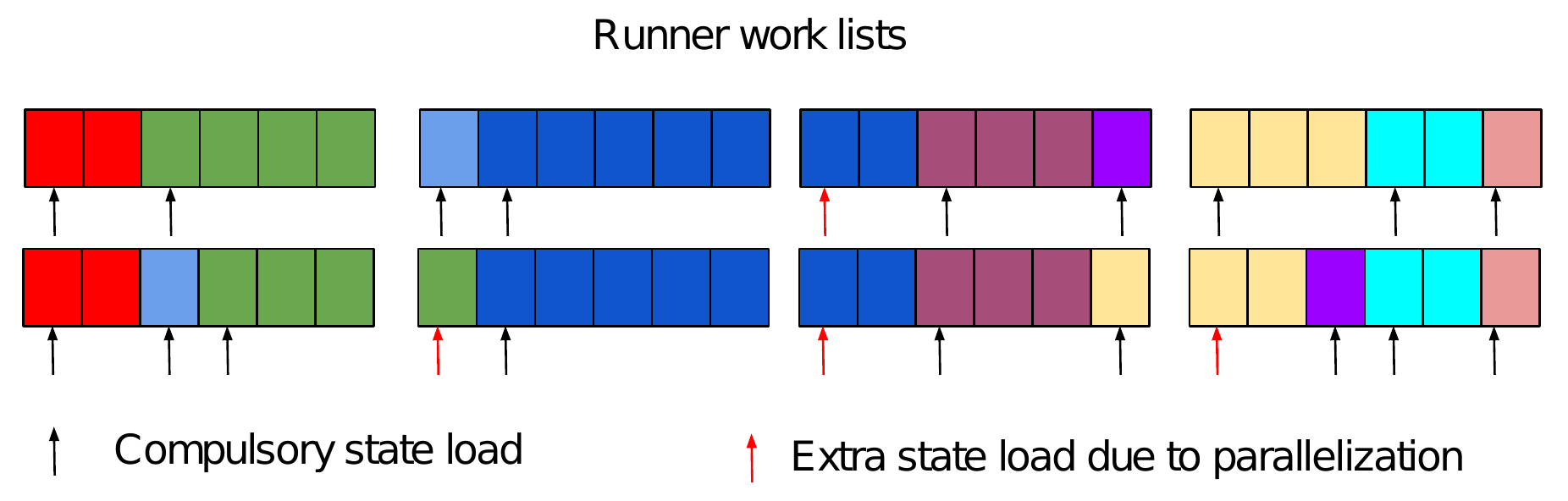}
    \caption{Two possible  schedules of 24 propagation tasks of equal duration on 4 runners.  All particles propagated
    from the same parent particle state have the same color (9 parents here). Top schedule is optimal
    with 9 compulsory  loads (one per parent), and one for the dark blue parent that cannot fit in
    one runner. The bottom schedule, with 2 more sate loads,  is a possible one that our on-line
    scheduling algorithm   can produce.   This is not optimal but still below the general $Q+R-1$
    bound    as the algorithm ensures that no more than $R-1$ "color cuts'' occur and avoids the
    same runner loads more than once a given parent particle state.}
    \label{fig:scheduling-algo}
\end{figure}

In this section, we present the scheduling algorithm implemented on the server to distribute the particle propagations to the runners. The algorithm aims to ensure an even load balancing between runners and minimizing the global particle loads, i.e. transfers of particles from global to local storage.

\subsubsection{Static Scheduling} \label{sec:static-schedule}

Let $R$ be the number of runners.  Let $P$ be the
number of particles to propagate. Resampling may lead to have some parent particles
drawn to be propagated several times. Let $Q$ the number of parent particles $q$, and $\alpha_q$ the number of times the  particle $q$ needs to be propagated. The total number of  particles to propagate is:

\begin{equation}
  P = \sum_{0 \leq q < Q}\alpha_q
\end{equation}

To assess the performance of our scheduling algorithm, we first derive a
lower and upper bound of the minimum number of particle loads $c^{*}$ for the static case, where:
\begin{inparaenum}[(i)]
  \item runners do not cache states,
  \item  the number of runners is constant, and
  \item  all particle propagations take the same amount of time.
\end{inparaenum}
Under these conditions, each runner propagates $\frac{P}{R}$
particles.  Without local cache, each parent particle $q$ needs to
be loaded at least once.  Therefore, the number of compulsory particle
loads is $Q$.  If $\alpha_q=1$ for all $q$, that is, every particle
is drawn only once, then $c^{*}=P$.  Otherwise,  parallelizing the
propagation on $R$ runners may require some particles to be loaded by more
than one runner, accounting for extra particle loads beyond
the compulsory ones.  Indeed,  each particle $q$ needs to be provided at
least on $s_{q}$ runners, where
\begin{equation} \label{eq:split}
    s_{q}=\ceil[\Bigg]{\frac{\alpha_{q}}{\frac{P}{R}}}.
\end{equation}
Distributed to $R$ runners, the list of $P$ particles is cut $R-1$ times.
Consequently, the extra particle loads are
at most $R-1$. This is visualized in~\Cref{fig:scheduling-algo}.
This upper bound occurs if all  particles are propagated from a single parent($Q=1$).
Thus, the minimum  number of particle loads  is tightly bound by
\begin{equation}
  Q \leq c^* \leq Q+R-1.\label{eq:boundary}
\end{equation}


We can apply  a static schedule that
respects the upper bound: distribute $\frac{P}{R}$ particles per
runner, where each parent particle $q$ is given
to no more than $s_q$ runners, and by imposing that runners do
not switch to the next particle before completing all
propagations associated to the current one.

\subsubsection{Dynamics List Scheduling} \label{sec:list-schedule}

However, we target a more general case.  We  soften the initial assumptions
now considering that  the number of runners can vary, and the time to propagate particles
may vary significantly and is not known beforehand (but we still have no cache).
In this context we propose to rely on  the classical  dynamic list~scheduling algorithm:
when  idle, a runner requests work from the server that returns a particle-id to propagate.
In the general case the  list~scheduling algorithm guarantees to  be at worst twice as long as the optimal schedule that requires to know the particle propagation time in advance~\cite{graham_bounds_1966, shmoys_scheduling_1991}.
Instead of blindly selecting  the next particle to propagate,  we adapt the static scheduling
strategy for particle selection with the goal of limiting the number of particle loads.
The scheduling is based on the
split factor~$s_{q}$ (\Cref{eq:split}). However, we adapt the static  scheduling to the
dynamic case by recomputing $s_{q}$ each time with the updated values of
$\alpha_{q}$, $P$, and $R$. To implement this algorithm on the server,
we need a bookkeeping of  the number of runners  $R_q$ currently
propagating particle $q$, and  the number $\alpha_q$  of remaining propagations
for particle $q$. Let $r$ be
the runner requesting a particle for propagation, the particle
distribution algorithm works as follows:
\begin{enumerate}[1:]
\item  If  $\alpha_{q} > 0$ for particle $q$ last propagated by $r$,
  decrement $\alpha_q$ and assign $q$ again. If $\alpha_{q} =0$ continue
  with (2);
\item  Select a different particle $q'$ with $\alpha_{q'} > 0$;
\item Compute split factor $s_{q'}$. If $R_{q'} < s_{q'}$ assign $q'$,
  increment $R_{q'}$, and decrement $\alpha_{q'}$. If $R_{q'} = s_{q'}$ continue with (2).
\end{enumerate}
Notice that when the server recognizes the loss of one runner, it needs to update the bookkeeping to
reintegrate the particle that this runner was propagating.

In conditions of even propagation time and a static number of runners, this algorithm leads to the same
distribution as for the static schedule and respects the upper bound of \Cref{eq:boundary}.

\subsubsection{Cache Aware Scheduling} \label{sec:cache-schedule}

We now remove the last assumption to propose a scheduling strategy that takes into consideration the
particle cache.  This is a heuristic build upon the  previous strategy and validated  though several experiments.  The particle selection strategy is:

\begin{enumerate}
\item Select a parent particle $p_i$ already loaded in the runner cache (cache hit); 
 \item Select a parent particle $p_i$ that is in no runner cache  (cache miss); 

 \item Select a particle $p_i$  fulfilling the split factor criterion  (cache miss); 

 \item  Select a parent particle $p_i$ with maximal split factor $s_i$ (even if voids  the split factor) (cache miss). 
 \end{enumerate}

The three first items comply with the scheduling  proposed in~\Cref{sec:list-schedule}. The first item  gives priority to particles already in the cache, before they may be evicted to provide space for a particle load.  The next two items  pursue with the strategy of~\Cref{sec:list-schedule},  favoring  particles  with no  previous propagation.  The rational is to start as soon as possible with new parent particles and, once in a cache, propagate them has often as required, and intend to  reduce the need for splitting. The last item departs from the strategy of ~\Cref{sec:list-schedule}, but its addition proved efficient by our experiments.  This case occurs when reaching the end of a cycle. It proved to be an efficient  strategy to keep runners busy, even at the cost of extra loads, to improve load balancing and so completion time.

\subsection{Job Submission and Monitoring}\label{sec:implementation:launcher}

The workflow is controlled by the \textbf{launcher}. The
launcher is the user entry point to configure and start the application.
The launcher starts first and is responsible to start and monitor the
runner and server instances, that all run in separate executables/jobs.
The launcher is also  in charge of killing and restarting the runners or server as requested
by the fault tolerance protocol (\Cref{sec:fautltolerance}), or for elasticity purpose.

The launcher tightly  interacts with the job scheduler (Slurm or OAR for instance) of the machine.
The launcher can be configured to submit one job per runner and server to the batch scheduler.  This
strategy offers the maximum flexibility for the batch scheduler to optimize the machine
ressource allocation, but the execution progress becomes very dependent on the machine availability.
The user  may need  more control on  the number of  concurrently running  runners. In that  case the
launcher can be set to request to the batch scheduler one or several large resource allocations  and
fit several runner instances in each one. To support this feature the launcher relies on
a combination of Slurm salloc/srun~\cite{SLURM-website}, or  OAR containers\cite{OAR-website}. For  even more flexible schemes, we plan to support workflow pilot-based  schedulers like Radical-Pilot~\cite{Radical-Pilot:2021} or QCG-PilotJob~\cite{QCG-PilotJob-website}.

\subsection{Fault Tolerance}\label{sec:fautltolerance}

The fault tolerance relies on the fast identification
of failures from runner and server instances. Runner failures are
detected in two different ways. Runner crashes are recognized by the
launcher, which is monitoring their execution using the cluster
scheduler.  Unresponsive runners are identified by the server relying on
timeouts for the particle propagations.  If propagations exceed the
timeout, the server requests the launcher to terminate the respective
runner. In both cases, the launcher eventually starts a new runner
instance. The new runner connects to the server and requests work.
If a runner fails, the server cancels the on-going propagation,
and the time spent in the propagation plus the time to recognize the
runner failure is lost.

Server failures are detected similarly, either directly if the server
crashes, or if the server exceeds a timeout. The
timeout is mediated by a periodical exchange of signals between launcher and server
(heartbeats). If the server fails, the launcher
terminates all runner instances and restarts the framework as a whole.
The server frequently stores the status of the propagations in
checkpoints, and in case of failures, the framework can recover to the
point of the last successful propagations.

Finally, a launcher failure is detected by the server monitoring the
heartbeat connection between launcher and server. In case of a missing
heartbeat, the server checkpoints the current particle state and shuts
down. In parallel,  the runners detect the server crash and shut
down, again using timeouts. While runner or server failures lead to an automatic restart, the framework needs to
be restarted manually if the launcher fails.

\subsection{Implementation Details}\label{sec:implementation:details}

The launcher and server are developed in Python.  The runner relies on the  simulation
code  instrumented  with our framework  API, supporting  C/C++, Fortran and Python.
The implementation reuses some  software components, like the launcher, from the framework developed
for \ac{enkf}  DA~\cite{friedemann-melissaDA:2022}. The distributed cache implementation
relies on the Fault Tolerance Interface (FTI)~\cite{bautista-gomez_fti_2011}. FTI is a multilevel
checkpoint-restart library supporting asynchronous checkpointing to global storage.
One of the main   modification performed to  FTI is related to its event loop.
Events are triggered in form of MPI communication between the application and FTI
processes. The events are identified by tags. To extend this
mechanism, we enabled  to register a  callback
function. This callback function is called inside the
event loop and can trigger user defined events using unique tags.
With this, it becomes possible to use the application checkpointing into all
available levels of reliability  FTI provides, and  to implement the
cache mangement  on the dedicated FTI processes.

The communication between helper  and model processes relies on asynchronous MPI messages.
Communications with the server are implemented in two steps for efficiency purpose.
Only rank~0 (master) of the application (i.e., model) communicator and the rank~0 (master) of the  helper process  communicator
communicate with the server. As a dynamic connection is needed,  each master connects to the server using a socket through the  ZMQ  library.
Information that needs to be propagated between  helper  or model processes relies on MPI collective communications in the associated  communicator (\Cref{fig:components:runner-server}).

The framework code is available at \url{https://gitlab.inria.fr/melissa/melissa-da}.

 \section{Experiments}\label{sec:exp}

\subsection{WRF Use Case}\label{sec:wrf_usecase}

\hide{
and lowering the difficulty of running the simulation across different computing platforms. However,
the complexity of atmospheric modelling does not lower its difficulty in performing skillful
prediction in climate modelling and weather forecasting.

The current WRF codes can support different simulation scales from large
eddy simulations (100~m in horizontal resolution), to tropical cyclone
(15~km), and to global domain (625~km in x- and 556~km in y-direction).
Benchmark tests on the AMD Epyc 7601 CPU for \uu{12}{km} and 2.5~km resolution cases over
Continental U.S. domain (CONUS12km, CONUS2.5km) show nearly linear scaling~\cite{Kashyap2019}.

We will experiment with WRF using a European domain (15~km horizontal resolution) based on the ERA5 data set. Up to 511 model instances (each parallelized on 39~compute cores of the Jean-Zay super computer) will run 2555 different particles to perform assimilation of cloud fraction observations for a day-ahead weather forecasting simulationw.
}

\begin{figure}[htb]
  \centering
  \includegraphics[width=0.6\linewidth]{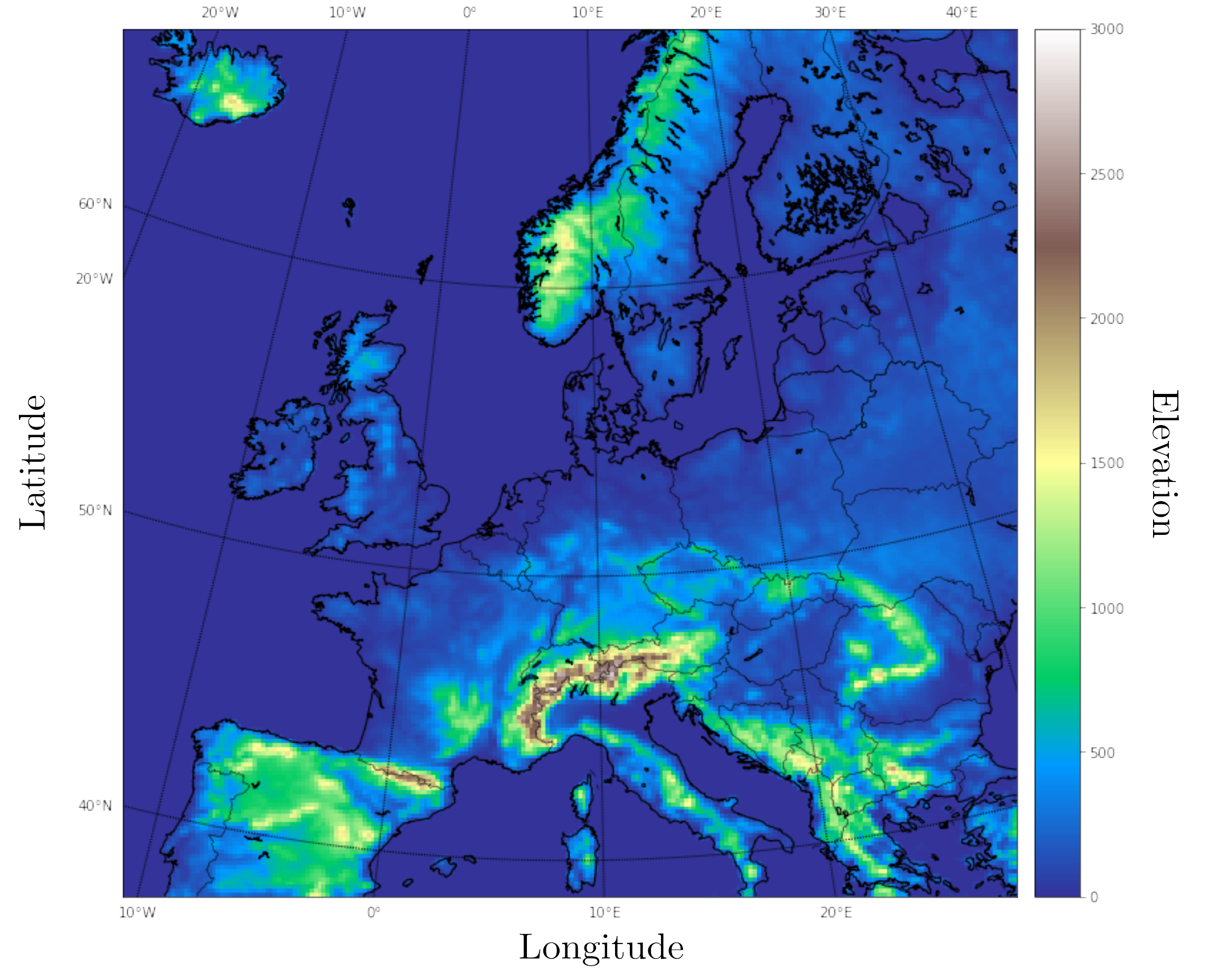}
    \caption{The topography of the target domain of Europe for the simulation.}
    \label{Figure:WRF15KMGEO}
\end{figure}

Experiments rely on an established \ac{nwp} system; the
\ac{wrf} (V3.7.1)\cite{Skamarock2008}. The core of \ac{wrf} is based on solving fully
compressible non-hydrostatic equations with complete Coriolis and
curvature terms, and a large set of physics options.  The simulation
domain covers most of Europe (See \Cref{Figure:WRF15KMGEO}) and is composed of
220 by 220 grid cells with a horizontal resolution of \uu{15}{km} and 49
vertical levels with uneven thickness. We perform one day-ahead weather
forecasting (24 hours of initial time plus 48 hours of production time) of an arbitrary date (2018-10-12) with 24-seconds or 100-seconds time steps.  The model employs the WSM6 microphysics, MYNN2 boundary layer physics, Grell-3 cumulus parameterization, Eta
Monin-Obukhov similarity surface layer processes, and RUC land surface
model. Also non-hydrostatics are activated to provide more details in simulated
clouds and precipitation. The input, initial, and boundary conditions are
based on the reanalyzed ERA5 dataset from the European Center for
Medium-Range Weather Forecasts (ECMWF), which is updated every 3 hours.
Data assimilation is performed using the cloud fraction (CFRACT). The particle weights are determined by comparison against the observed cloud mask obtained from the EUMETSAT Level-2 satellite data of the cloud mask. The simulated cloud fraction is converted into cloud mask and the observed cloud mask data is upscaling to the size of the model gridcells for the further applications. 
The data is hourly available, thus, one assimilation cycle comprises 150 (36)
model time steps (150~$\times$~\uu{24}{s}~$\widehat{=}$~\uu{1}{h} or 36~$\times$~\uu{100}{s}~$\widehat{=}$~\uu{1}{h}).

\hide{
  Europe climate is classified as temperature and continental climate conditions for the most part of west and east, respectively \cite{Beck2018}. With the increase of extreme weather events, a more skilled weather prediction is more urgent and needed. In the following we rely on the WRF model to assimilate a study case for energy meteorology simulations that provide meteorological input for the calculation on photovoltaic and wind power generation. Simulation results are also used as training data for future machine learning models predicting photovoltaic and wind power generation. \sebastian{cite needed here!}
The used domain covers most Europe. There are 220 by 220 grid cells with horizontal resolution of 15~km and 49 vertical levels with uneven thickness. The topography of the model domain is shown in \Cref{Figure:WRF15KMGEO}. Such model setups are used for a day-ahead weather forecast. The input dataset is based on the reanalysed ERA5 dataset from the European Centre for Medium--Range Weather Forecasts (ECMWF). We randomly choose 2018-07-19 to simulate 48 hours for time steps of 100 seconds. The model employs the WSM6 microphysics, MYNN2 boundary layer physics, Grell-3 cumulus parameterization, Eta Monin-Obukhov similarity surface layer processes, and RUC land surface model. We also employ non-hydrostatics to have more detail in simulated cloud and precipitation. Data assimilation is performed regarding the cloud cover fraction (CFRACT) variable. Each particle's cloud cover is compared with satellite data, which is the cloud fractional cover data from EUMETSAT CMSAF product \cite{Stengel2014},
and a weight for resampling is calculated accordingly. The data is available hourly so we decided to resample every 36 model time steps ($36 \times 100 \text{s } \widehat{=}\; 1 \text{h}$) to assimilate all observation data available and to especially stress the architecture.
}

The experiments presented in this article
leverage our proposed \ac{pf} implementation with a sample size of
up to \numprint{2555}~particles on the European domain. In that case, we
utilize \numprint{20442}~compute cores on 512~Nodes of the Jean-Zay
supercomputer. The compute nodes are equipped with 2~Intel Cascade Lake
\numprint{6248}~processors, summing  up to 40~cores with \uu{2.5}{GHz}
and \uu{192}{GiB}~RAM per node. Intel Omni-Path (100~GB/s) connects the
compute nodes, and the global file system is an IBM Spectrum Scale
(ex-GPFS) parallel file system with SSD disks (GridScaler GS18K~SSD).
For all experiments the node-local caches were stored on RAM disk.
In~\Cref{tab:overview} we list the parameters of our main experiments. 

The meteorological state of the European domain associated to
one particle comprises \uu{2.5}{GiB} of  data.  Hence, the
data from \numprint{2555} particles for the full simulation period of
48h (48 time steps) correspond to an aggregated size of about
\uu{300}{TiB}. The experiments performed for this article,
including small beta-stage experiments, account for about \numprint{900 000}~CPU
hours split between the JUWELS, Jean-Zay and  Marenostrum
supercomputers.

\hide{

The experiments performed for this article,
including small beta-stage experiments, account for about \numprint{220000}~CPU
hours split between the JUWELS, Jean-Zay and  Marenostrum
supercomputers.

  Jean-zay:
   - XXX
   Juwel
   - Sebastian: XXX
      - got 'some' hours on Juwels - its about 78k coreh for the last experiments only (Those I could figure out on the judoor platform). At least three times this value was used for all the experiments but ATM I cannot access ligone from where I used to access jean-zay and juwels so I cannot figure out more precise information. In total for the PhD (till the defense date only) about 960k coreh were used. At least 216k were for the other paper. But I will try to get access again to the supercomputer adding another gateway IP so I can get in there again
   - Yen-Sen: 190 000 core-hours
       - Sorry to raise this issue this late, but does the 220,000 core-h includes mine? I've spent 189.98K for this task. Don't know if you want to include. 
   -Kai: XXXX

   Marenostrum
   - Kai: XXXX
   }



\begin{table}[htb]
    \hspace{-0.4cm}
   \footnotesize
   \centering
  \begin{tabularx}{\linewidth}{lXXXX}
\toprule
  \multicolumn{5}{c}{Experimental Setup} \\
\midrule
  Particles &   \numprint{315}  &   \numprint{635}  &   \numprint{1275}  &   \numprint{2555} \\
\makecell[l]{Number of runners}        &     \numprint{63} &     \numprint{127} &     \numprint{255} &     \numprint{511} \\
\makecell[l]{Number of nodes}         &     \numprint{64} &     \numprint{128} &     \numprint{256} &     \numprint{512} \\
\makecell[l]{Model processes}      &      \numprint{2457} &      \numprint{4953} &      \numprint{9945} &      \numprint{19929} \\
\makecell[l]{Particles per runner (avg.)}            &   \numprint{5} &   \numprint{5} &   \numprint{5} &   \numprint{5} \\

\makecell[l]{Particle state size (GiB)}        & \numprint{2.5} & \numprint{2.5} & \numprint{2.5} & \numprint{2.5} \\
\midrule
\multicolumn{5}{c}{Performance Data} \\
\midrule
\makecell[l]{Scaling efficiency}                                          &    $92\%$ &    $91\%$ &    $92\%$  &    $87\%$ \\
      \makecell[l]{Resampling (ms)}                                                                    &    \numprint{2.21} &   \numprint{4.06} &   \numprint{8.16} &   \numprint{16.37} \\
  \makecell[l]{Assimilation cycle (s)}                                                               &   \numprint{136} &   \numprint{138} &   \numprint{139} &   \numprint{146} \\
  \makecell[l]{Propagation (s)}                                    &   \numprint{25.1} &   \numprint{25.2} &   \numprint{25.1} &   \numprint{25.0} \\
  \makecell[l]{Load particle state\\from PFS to cache (s)}                                       &    \numprint{2.1} &    \numprint{2.1} &    \numprint{2.4} &    \numprint{4.1} \\
  \makecell[l]{Write particle state\\from cache to PFS (s)}                                       &    \numprint{1.4} &    \numprint{1.6} &    \numprint{1.8} &    \numprint{2.3} \\
    \makecell[l]{Writes to PFS per cycle (TiB)}              &   \numprint{0.77} &   \numprint{1.55} &   \numprint{3.11} &   \numprint{6.24} \\
    \makecell[l]{Reads from  PFS per cycle (TiB)}                                                & \mbox{\numprint{0.30}-\numprint{0.40}} &   \mbox{\numprint{0.64}-\numprint{0.79}} &   \mbox{\numprint{1.27}-\numprint{1.79}} &    \mbox{\numprint{2.54}-\numprint{3.82}} \\
\bottomrule
\end{tabularx}

    \caption{Experimental setting and performance overview at 4 different scales. The times are given as average in all cases. Model time steps were set to 100 seconds.}
    \label{tab:overview}
  \end{table}

\subsection{Runner Activity}\label{sec:overlapexp}



\begin{figure}
    \centering
    \includegraphics[width=.9\linewidth]{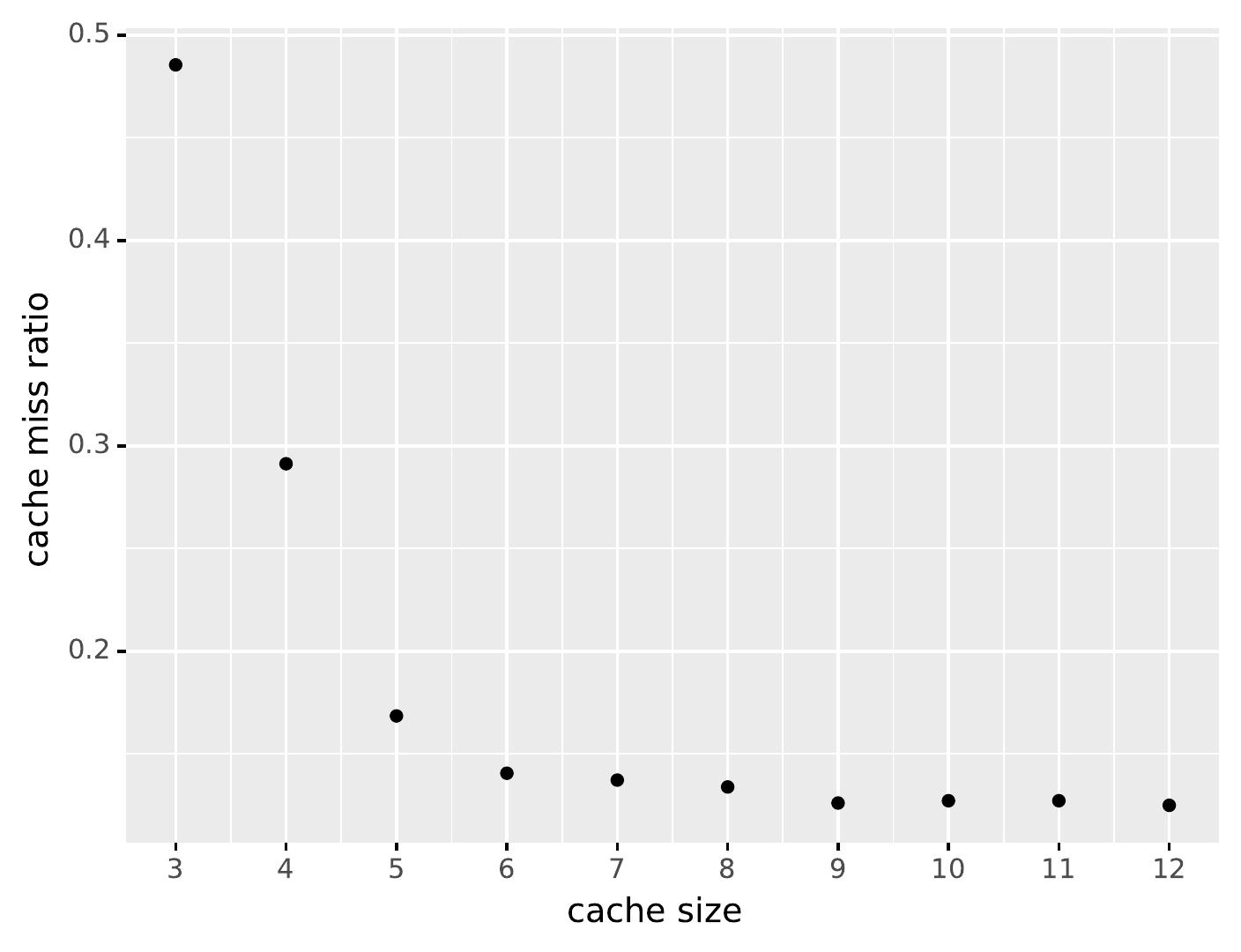}
    \caption{Cache miss ratio for different cache sizes on each runner.
      In total 128 particles  run on 32 runners. 
        First and last assimilation cycles were disregarded to remove warm~up effects and not
    fully recorded cycles.}
    \label{fig:cachemissratio}
\end{figure}
\begin{figure}[htb]
  \centering
  \includegraphics[width=0.9\linewidth]{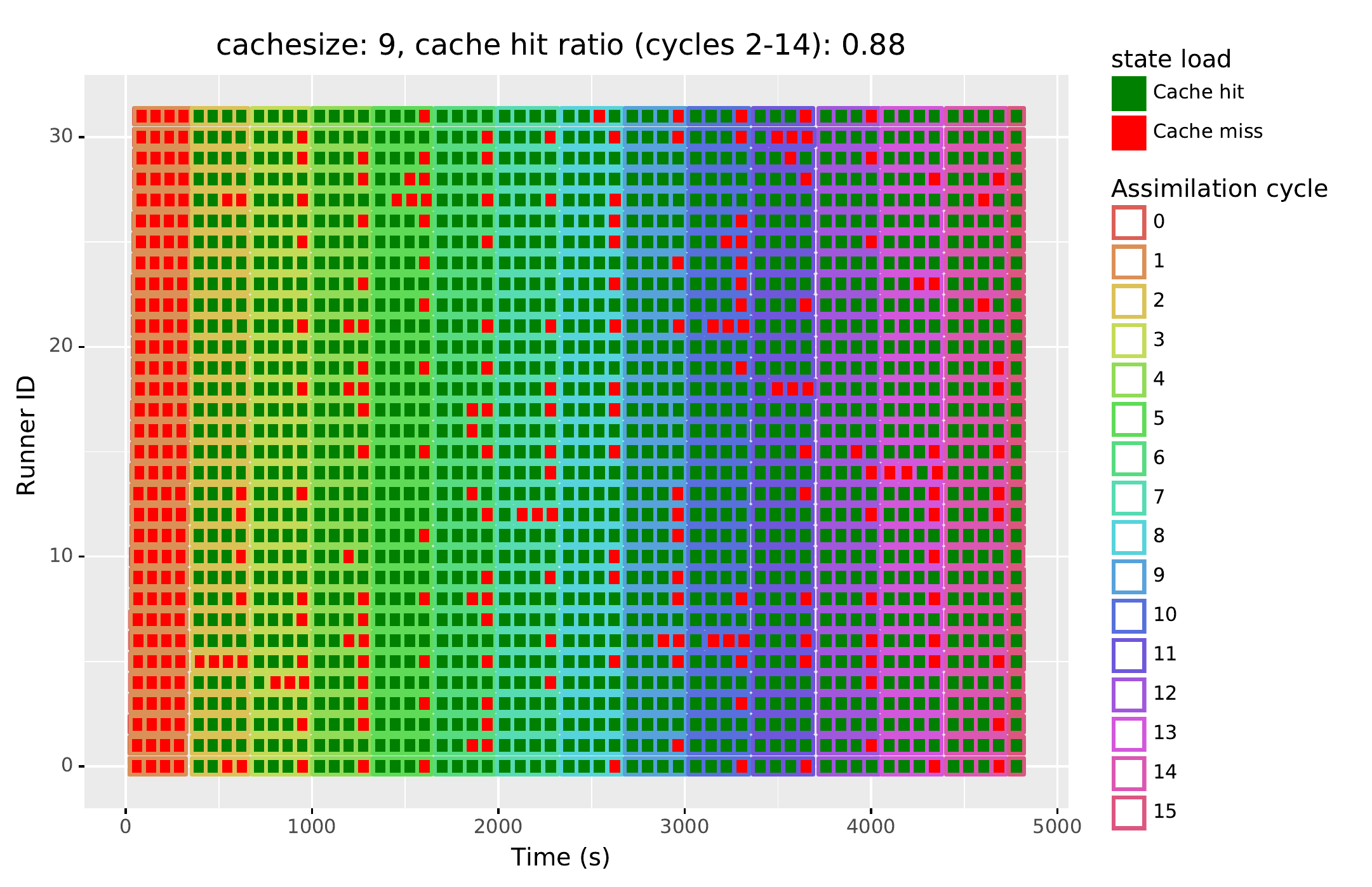}
    \caption{Gantt chart of  particle propagations executed by the 32
      runners over 15 assimilation cycles.  Tasks are green if the associated parent particle state was already present in the runner cache and did not require a load from the PFS (red otherwise).  
}
\label{fig:many_runners_trace}
\end{figure}

The benefit of the local cache in combination with the cache-aware
scheduling leads to a drastic reduction in transfers from global to
local file system layers.
The cache hit ratio, i.e., the ratio of particles found
 inside the cache to the total number of particle loads, 
 depends on the cache size and the ratio of particles per runner.
\Cref{fig:cachemissratio} shows how the cache misses develop for
different cache sizes.
Our experiments demonstrate a cache hit ratio of
88~\% for \numprint{128} particles, 32 runners, and a cache size
of 9 particles. This translates to a saving of 88~\% in transfers
from global to local storage. The pattern of cache hits and misses is
visualized in~\Cref{fig:many_runners_trace}.
The initial phase is dominated by starting
up the runners, and all the particles are fetched from the global
storage~(cache warm~up). But beginning with the next assimilation cycle, the low
transfer ratio from global to local storage starts to establish.

\begin{figure}[htb]
  \centering
\includegraphics[width=0.9\linewidth]{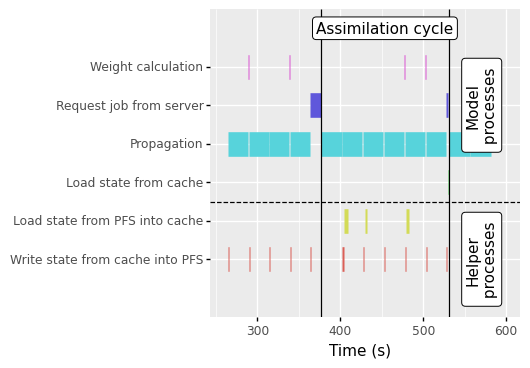}
\caption{Trace detailing the activity of  a runner over the course of an assimilation cycle. Helper processes enable to keep model processes busy  with particle propagation, except at the end of assimilation cycles when
  they wait for the server to finish particle resampling (dark blue). Some activities are so thin that they are not visible here (state copies from cache to model).
}
\label{fig:trace}
\end{figure}

Runners are  designed to separate I/O operations to the PFS   from 
other tasks: model processes only perform local I/O operations. 
We observe in our experiments that
this leads to a high computational efficiency. The local I/O accesses are negligible compared to the
computational tasks ($< 0.1$ s compared to up to \uu{6}{s}). Some general idle
periods can be observed between assimilation cycles when runners
are waiting for the last propagations of one cycle to finish so that the server can normalize weights, resample and start to
distribute work again. This is illustrated in~\Cref{fig:trace} where we
show a trace recorded from the execution of an arbitrary runner. The trace
 illustrates the efficiency of the runners in
performing the actual tasks of the simulation, particle propagation and
weight calculation, while the I/O tasks are moved to the background.

A global parallelization of the computational tasks is achieved by dynamically 
distributing the particle propagations to the available runners. The fully
parallelized case corresponds to $R=P$, i.e., the number of runners
matches the number of particles. The sequential case corresponds to
$R=1$, i.e., all propagations are performed by only one runner. However,
The best parallel efficiency is achieved at values between those
limits.  Because WRF  propagates particles 
 with very low time variability (maximum variation of 10\%),
we observe an even distribution of propagations to runners
when  $R$ divides $P$ (\Cref{fig:many_runners_trace}).
A single-particle propagation takes between 24 and 26.5 seconds, globally making from $87\%$ to  $92\%$ of
an average assimilation cycle. Calculating weights takes $1\%$ of the time and communicating
with the server  from $7\% $to  $12\%$  including the idle time at the end of each
cycle (\Cref{tab:overview} -- \textit{Performance Data}). The extra resources for helper
processes, one core per runner node, and the server, 1 node, comprise
only $2.7\%$ for our largest experiment at 512 nodes. On the other hand,
leveraging the runner's particle cache, and the cache aware dynamic
scheduling on the server, move $>97\%$ of the state loads completely into the background.
Loading and writing particle states synchronously would otherwise add
about $6.4$ seconds to each single-particle propagation corresponding to
$14\%$ of the average propagation time (\Cref{tab:overview} -- \textit{\numprint{2555} particles}).

Note that in contrast to the traditional offline approach, we start-up  the
numerical simulation code only once for several particle propagations.
The setup involves the request and allocation of the runner job and
initializing the simulation code. On the other hand, the traditional
offline approach associates each particle propagation with a different
 job, and the start-up has to be performed anew for each particle.
Starting up the WRF model on the European domain on one node until the
first model propagation begins takes 3-\uu{4}{s}, excluding the
provisioning of the job allocation via the cluster scheduler.





\subsection{Server Activity}

\begin{figure}[htb]
      \centering
\includegraphics[width=0.9\linewidth]{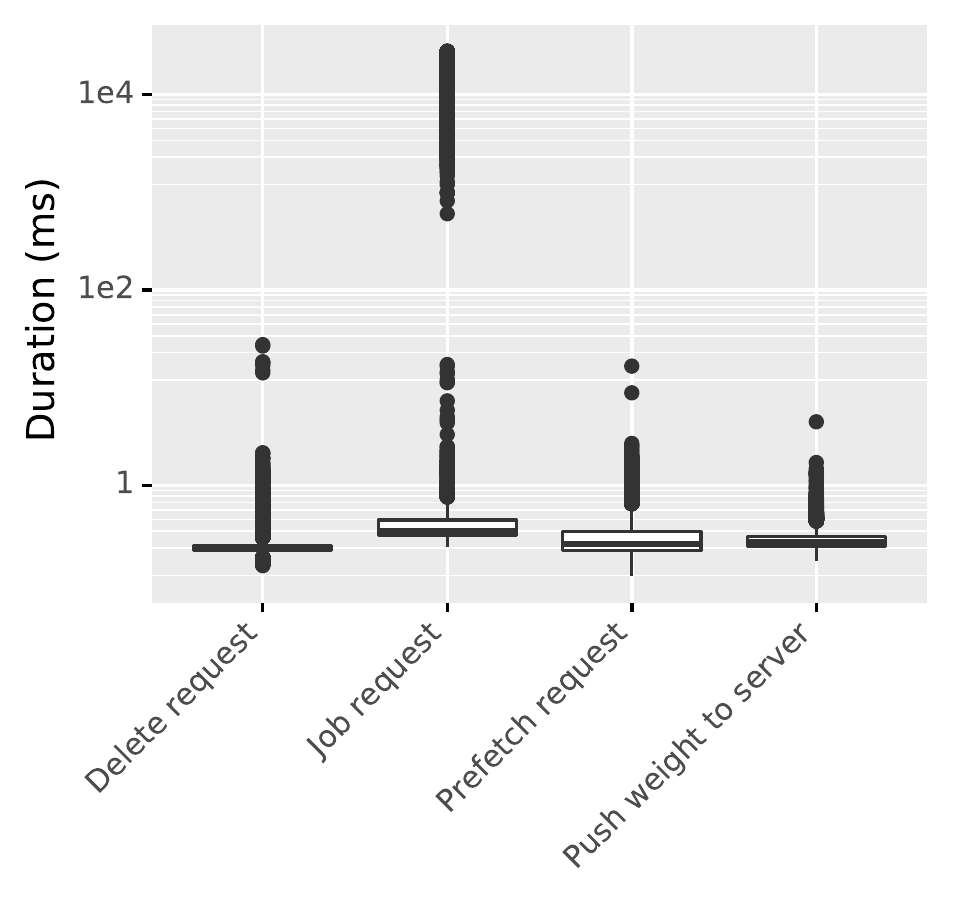}
  \caption{
  Server response times on runner requests.}
\label{fig:server_resp}
\end{figure}

We further measured the server responsivity to runner requests. The
response time is always in the order of a few hundred microseconds,
except for some job requests that take up to a few
seconds~(\Cref{fig:server_resp}).  However, these are outliers at the
end of the assimilation cycle, resulting from idle times due to the
load balancing and the particle filter update. During our largest
experiments with 511 runners, the server processes 676 requests per
second at maximum load. This shows that the server is fast enough to
support this scale, even though it is a sequential python code. Simple
optimizations are at reach if the server needs to be accelerated (e.g.,
adding parallelization).
\hide{ but it is assumed to be fast enough for the following orders of
magnitude in terms of runner and particle count.}

\subsection{State Transfers To/From PFS}

\hide{
Recall that we have $Q$ unique particles and a total number of $P$
particles with $P=\sum\,\alpha_q$. The alphas correspond to the
remaining propagations for the unique particles $q \in Q$\sebastian{@kai: check this please: is Q the number of particles or the set containing all the qs ($Q=\{q_1,q_2...\}$)}. For the
sake of load balancing, we need to propagate some unique particles on
multiple runners, hence, transfer those particles to additional runners besides
the one that generated the particle during the previous cycle.  The
cache-aware dynamic schyeduling algorithm minimizes the transfers from
global to local storage as we explained
in~\Cref{sec:implementation:scheduling}. Additionally, we implemented an
asynchronous prefetching mechanism for those particles that still need
to be transferred from global to local storage.
}

\begin{figure}
  \centering
\includegraphics[width=0.9\linewidth]{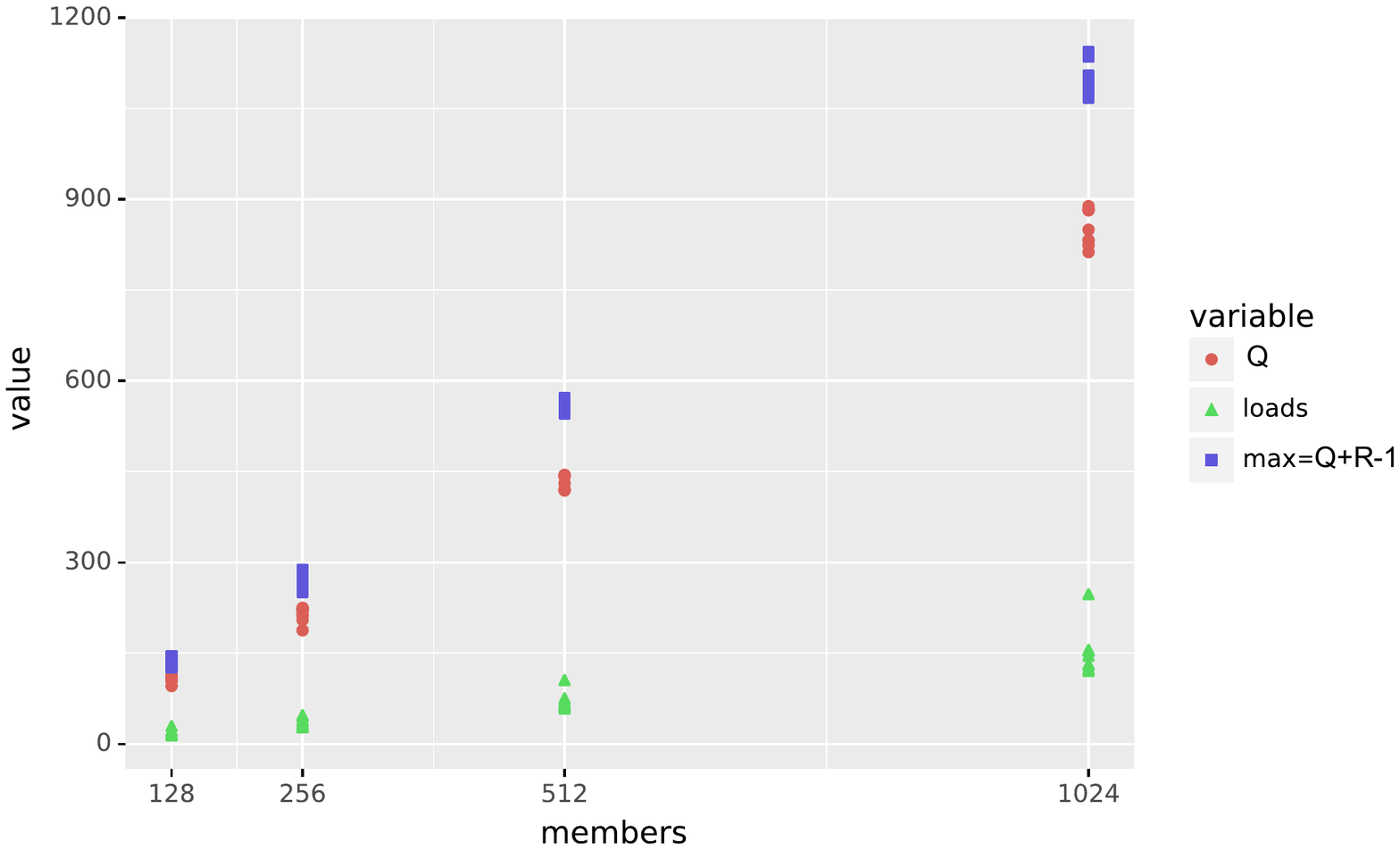}
\caption{Number of parent particles $Q$, particles loads from the PFS to the cache, and $Q+R-1$ upper bound from  \Cref{eq:boundary} for different ensemble sizes, a  cache size of~9~particles with~4~particles per runner.}\label{fig:undercutcriteria}
\end{figure}

Next, we take a closer look at the particle loads from the PFS (\Cref{fig:undercutcriteria}).
With a sample size of $1024$ particles, leveraging $256$ runners, and a local cache size of 9 particles, between \numprint{121}  and
\numprint{248} particles are loaded to the cache during each
cycle. The number of distinct  parent  particles $Q$ propagated per
cycle lies between \numprint{813} and \numprint{889}. Each one is propagated at most 
5 times to sum to a total of $1024$ particles. The cache enables to achieve significantly less loads than the
$Q+R-1$ upper bound expected with static scheduling and no cache (\Cref{eq:boundary}).

\begin{figure}[htb]
  \centering
\includegraphics[width=0.9\linewidth]{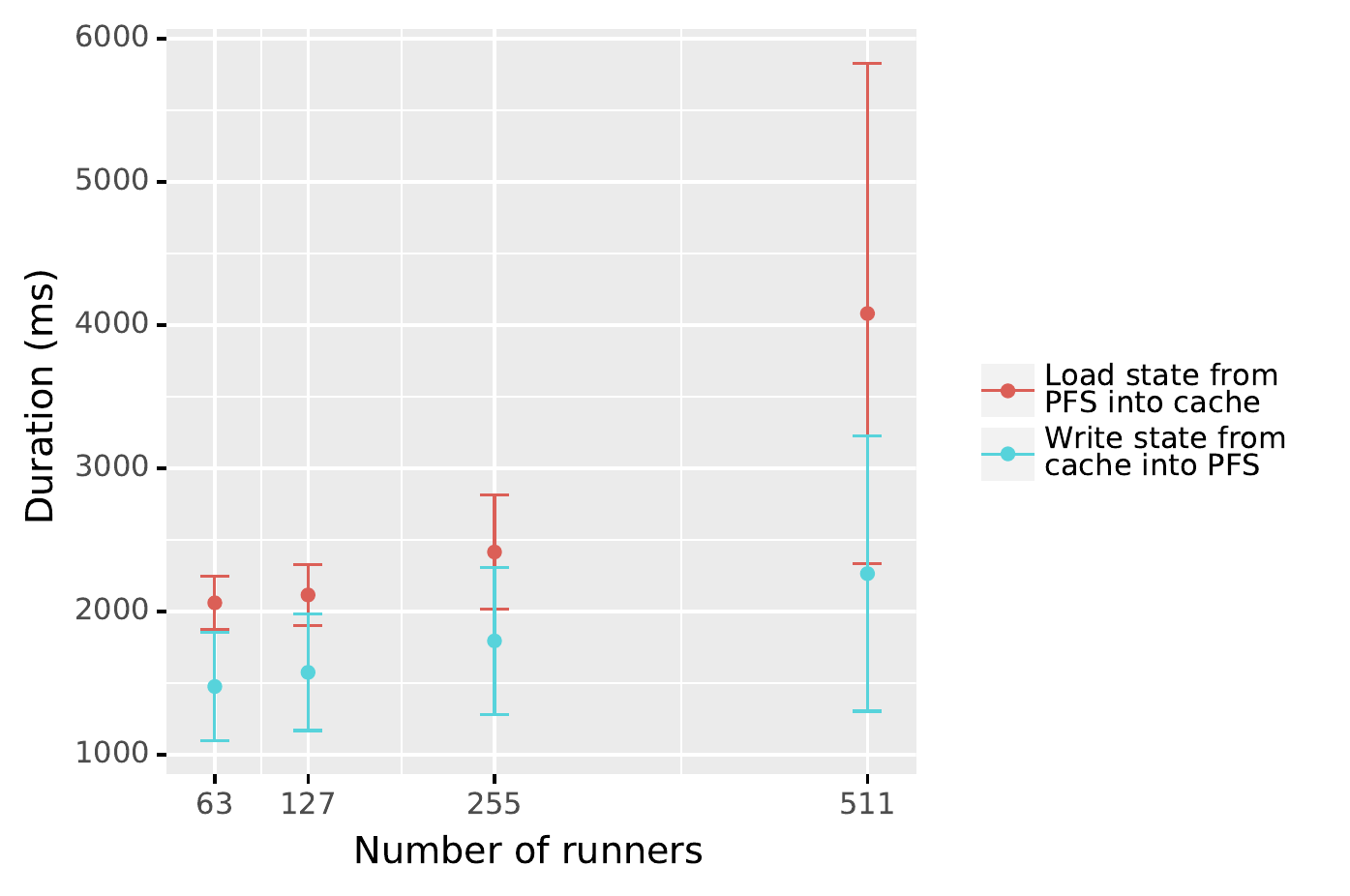}
\caption{Mean time  to load or store particle states of \uu{2.5}{GiB} from / to the PFS  with different numbers of runners. }\label{fig:pfs_times}
\end{figure}

\begin{figure}[htb]
  \centering
\includegraphics[width=0.9\linewidth]{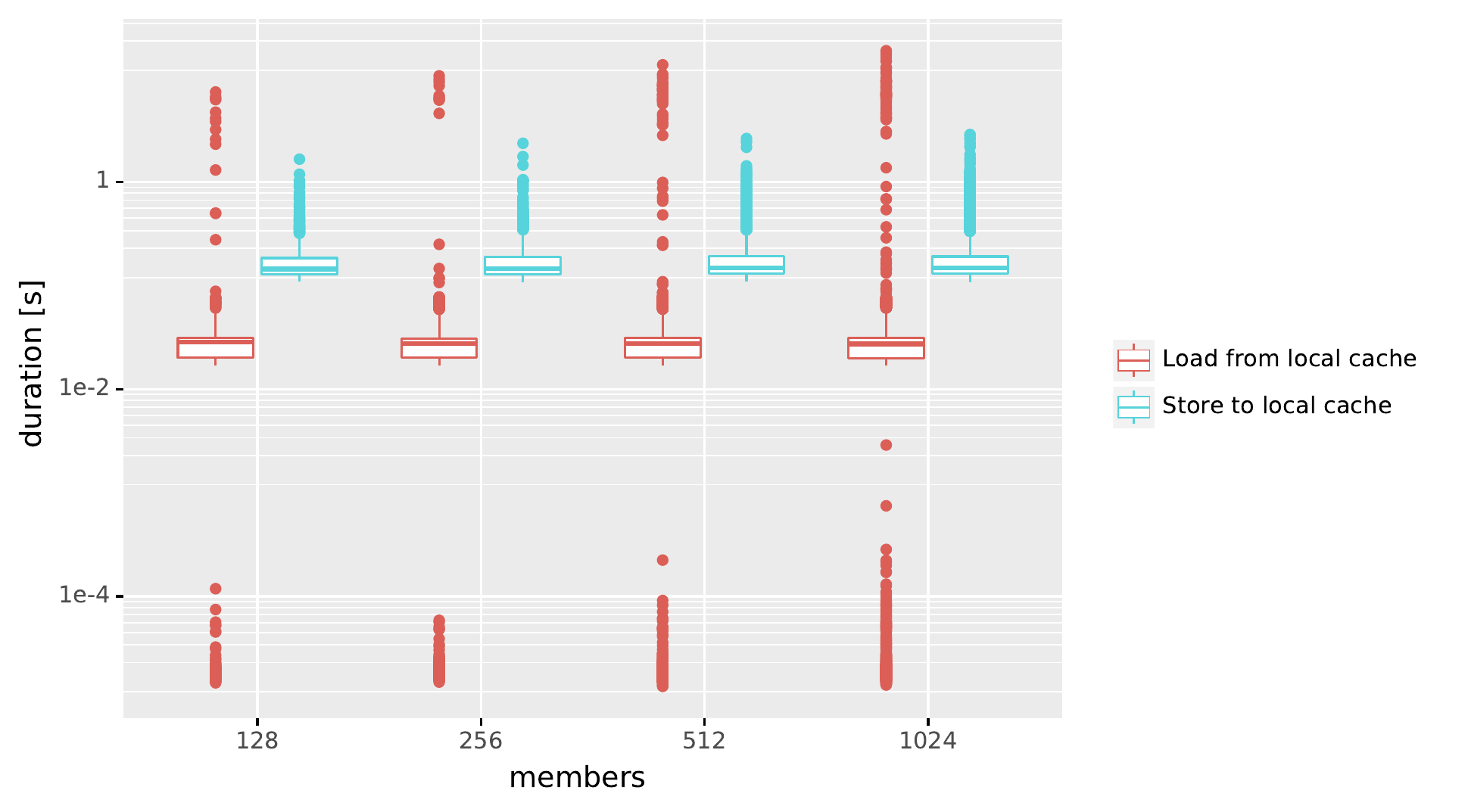}
\caption{Box plot of the time spent for  loads and  stores
from/to the local cache with different numbers of
particles.}\label{fig:localloadstore}
\end{figure}

The access times to the \ac{pfs} (load/store) vary
significantly and increase with the number of runners
(\Cref{fig:pfs_times}), showing that our application alone can stress
the \ac{pfs}~\footnote{these numbers may also be impacted by other jobs
on the cluster}.  Each particle is associated with \uu{2.5}{GiB} of
 data. During  each assimilation cycle, all the propagated particles are written
 to the \ac{pfs} for supporting fault-tolerance and  dynamic load balancing. 
 For our experiments at 512 nodes with \numprint{2555}
particles, this accumulates to about 6.2~TiB of data each cycle
(compare~\Cref{tab:overview}).  However,  our experiments on the
Jean-Zay and JUWELS supercomputer demonstrate that our framework
performs most of those transfers asynchronously (\Cref{sec:overlapexp}).
In less than 2\% of the cases, the model processes  wait more than 0.1~seconds for a particle to
be loaded corresponding to cases where helper processes  do not (entirely) finish prefetching.
Time to perform the local loads and stores from the  cache shows a  constant  average  independently on the number of runners  (\Cref{fig:localloadstore}).

\subsection{Fault Tolerance, Elasticity and Load Balancing} 
\begin{figure}[htb]
      \centering
\includegraphics[width=0.9\linewidth]{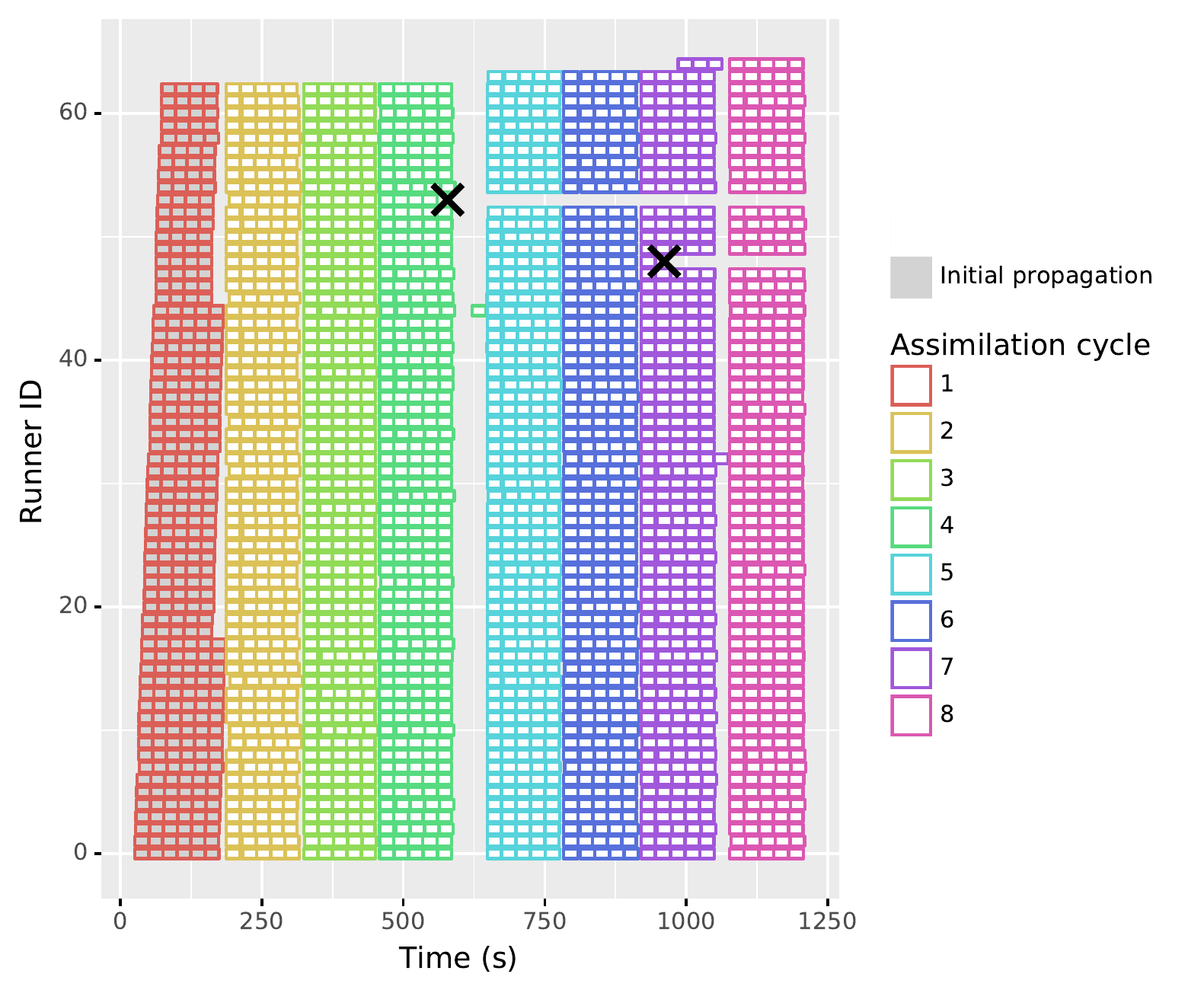}
  \caption{Gantt chart of particle propagations executed by the 63 runners over 8
assimilation cycles. After runners $\#48$ and $\#53$ crashed (black cross), two new ones restarted (top 2 runners $\#63$ and $\#64$).}
\label{fig:elasticity}
\end{figure}

\hide{
Fault tolerance relies on the accessibility of
the particles on persistent global storage. We have shown in the
last sections that we move all propagated particles to the \ac{pfs}
asynchronously  with  the helper processes. The second pillar of
fault tolerance  relies on its elasticity; The number of
runners can dynamically be adjusted. The fault tolerance protocol is the
same as in the original Melissa-DA framework (compare Friedemann et
al.~\cite{friedemann-melissaDA:2022} -- \textit{Fault tolerance and
elasticity}). If runners fail, they are
replaced by new runners, so that we guarantee a constant number of
runners at all times. As the particles are globally available, the new
runner can fetch any particle that is assigned to it for propagation.
}

Fault tolerance relies on 1) persisting the particle to the \ac{pfs}
2) the framework elasticity enabling to adjust dynamically  the number of
runners.  If a runner fails, the launcher requests the execution of a new one, so  as to maintain  a constant number of
runners. Once this new runner connects to the server, it asks for a particle to propagate
to the server,  assigned according to the load balancing algorithm.

We tested the fault tolerance and elasticity on an experiment with 63
runners provoking the crash of 2~runners (\Cref{fig:elasticity}).
First, notice that the fault tolerance algorithm reacts appropriately as
it restarts a new runner after each crash. The first crash (runner~$\#53$)
occurs in the worst-case situation: just when propagating the last
particle  of the current cycle, leading to a significant idle period.
The idle period is caused first, because the server needs to wait for
the timeout (set to \uu{60}{s}) to acknowledge that  runner~$\#53$ is
unresponsive and second, there is no work left except the particle that runner $\#53$ was propagating,  which is re-assigned to runner~$\#44$.
Meanwhile  all other runners stay idle until the beginning of the next cycle. If the crash happens earlier
during a cycle, smaller idle periods appear. This can be
observed during the second crash (runner~$\#48$), as the other runners are
kept busy with propagation work. \hide{The idle period at the end of the
affected cycle is merely a consequence of the dynamic load balancing
coming into play.} We generally observe an efficient load balancing, as
the work load is kept well distributed amongst runners, even when their
number varies.

\hide{
Similarly, the framework provides elasticity to
dynamically adapt the number of runners. The dynamic scheduling ensures
an efficient load balancing in either case.

In \autoref{fig:elasticity} we execute an assimilation as described above but with 63 runners (still 5 particles per runner) where we crash some runners to demonstrate fault tollerance and elasticity. Black crosses mark runners that were crashed.
The trace of model propagations of restarted runners are added to the top in the plot. Since runners are crashed via the appropriate command to the batch scheduler, the launcher detects the failure fast and starts up a new runner that registers at the server within 25~seconds. Still the server needs some time to detect that a scheduled job times out (in the worst case up to 60~seconds as set as runner timeout). Thus the failure of the job on runner 53 is detected by the server \emph{after} the new runner 63 was already started by the launcher and the missing job is rescheduled to runner 44 only after the timeout was hit.

The same mechanism enables the framework's elasticity permitting to change the amount of used resources during runtime. 

A changing allocation size impacts the execution time since more or less model propagations can be performed in parallel.
Since in our experiments the propagation times from particle to particle are very similar (less then 10\% relative fluctuation),
adding or removing only few new runners  the number of particles per runner introducing some load imbalance.

Model codes where propagation times vary, be it due to iterative solvers performing a different number of steps until a convergence level is reached, or due to different physics used per particle can be smoothly load balanced by the used list scheduling algorithm. This would also permit to execute efficiently when runners are started on heterogeneous resources, e.g., compute nodes leveraging GPU accelerators or not or when the particle count is changed on the fly to, e.g., comply with some convergence criteria. This is left to be explored in future.

}

\subsection{Scaling}\label{sec:scaling}

\begin{figure}[htb]
  \centering
  \includegraphics[width=0.6\linewidth]{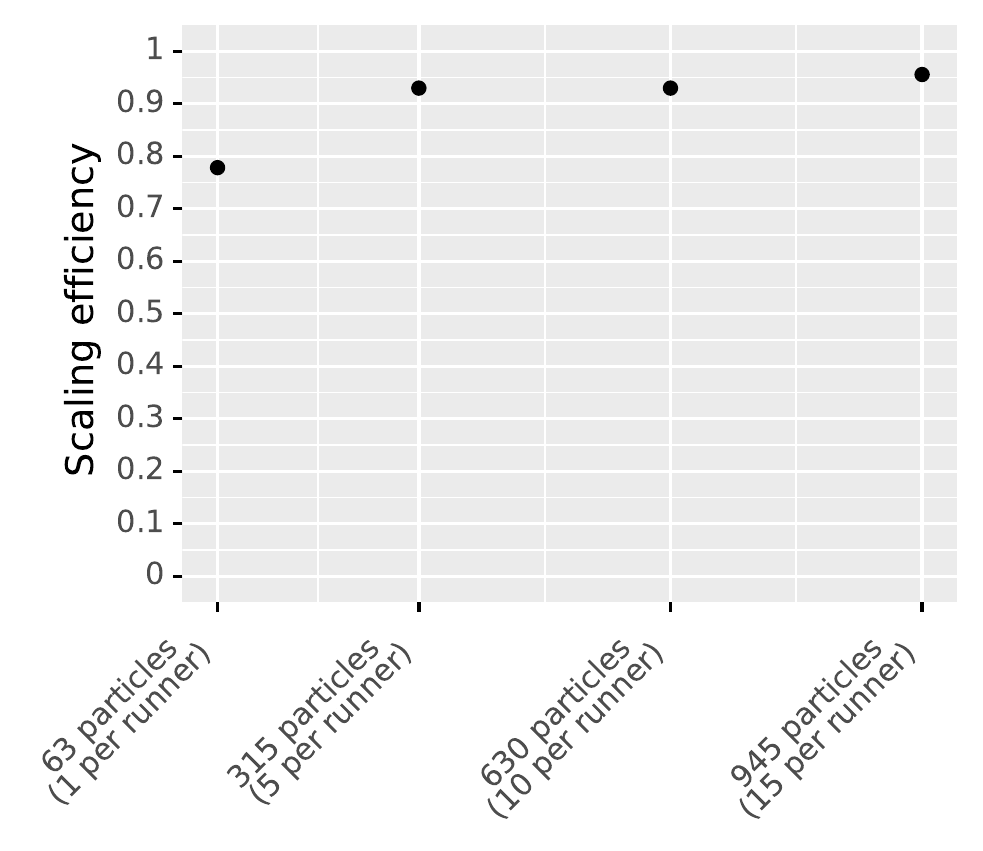}
    \caption{Scaling efficiency using different numbers of particles with  63~runners. One runner
      sets the reference case.
    }
\label{fig:strong_scaling}
\end{figure}

\begin{figure}[htb]
  \centering
  \includegraphics[width=0.6\linewidth]{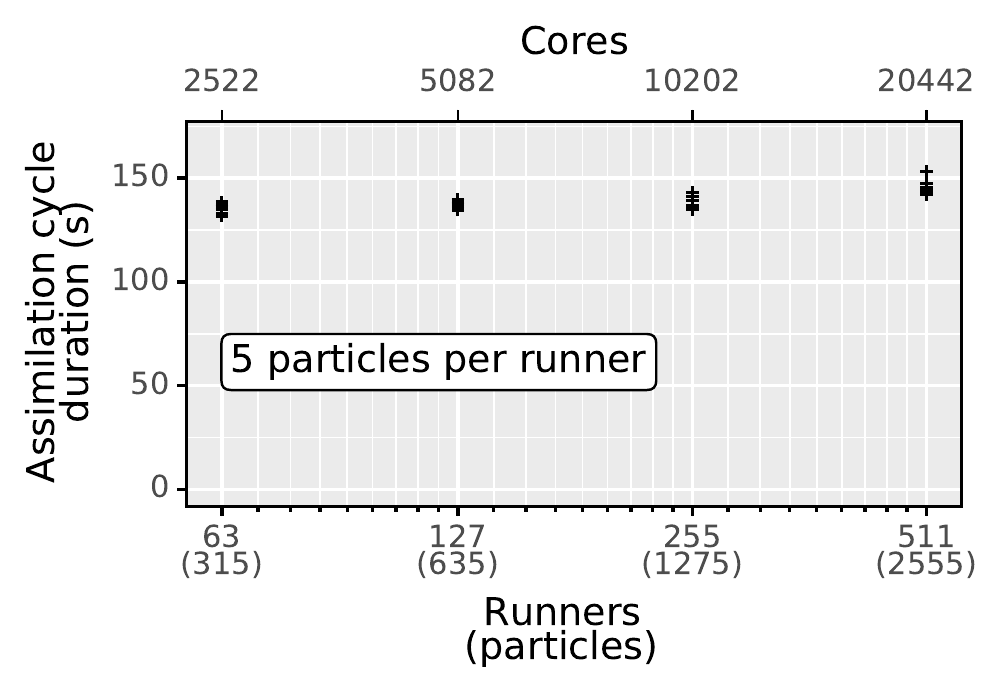}
\caption{Weak scaling  performance test: assimilation cycle duration for different numbers of runners, but always 5 particles per runner.
    }
\label{fig:weak_scaling}
\end{figure}

We evaluated the performance of the particle filter in a strong scaling
scenario, constant number of runners while increasing the number of
particles, and a weak scaling scenario, constant particle-to-runner
ratio while increasing the number of runners. In the strong scaling
scenario we observe that the runner utilization shows an upwards trend
when  increasing the number of particles per runner, with a plateau  at about 96\% (\Cref{fig:strong_scaling}).
As global I/O operations are almost completely shadowed, thanks to the asynchronous  prefetching,
increasing the number of particles per runner  mainly enables to better
amortize the cost of the synchronization associated with resampling.
We observe an almost constant time for the assimilation
cycle, demonstrating a desirable weak scaling behavior.  The time for the cycles increase
only by $8\%$ from $63$ to $511$ runners, indicating an efficient scaling behavior of the framework up to production scale
(\Cref{fig:weak_scaling}).
Particle filtering with WRF on a European domain for short-range
weather prediction at this scale is an important advancement of the
previous work done by Berndt et.\ al.\
\cite{berndt_predictability_2018}. Moreover, besides assimilating at a higher
frequency, our proposal offers fault tolerance, automatic
load balancing and elasticity while minimizing the I/O cost and time to calculate weights.

\hide{
 Berndt et. al.~\cite{berndt_predictability_2018} run WRF with a similar configuration with up to 4096 particles, on up to 262,144 cores. The different members are assembled in a a large MPI code. The code supports neither particle virtualization, nor fault tolerance or elasticity.  The experiment close to our scale at 2048 members runs one propagation cycle in 30 seconds  on  131 000 cores. The lack of particle virtualization requires to  reserve a very large number of cores, a significant fraction of the supercomputer, which is often  difficult. Particle virtualization enables us to run a similar number of particles on about 20k cores only. If we take the freedom to compare directly the compute times, we perform a cycle in  146 seconds, so about $1.3$ times faster.
}

\subsection{Comparison to a File-based Approach}

\begin{table}[!h]
  \centering
  \scriptsize  
  \begin{tabular}{|r|r|r|r|r|r|r|r|}
    \hline 
    &  \multicolumn{3}{c|}{Melissa} & \multicolumn{3}{c|}{ESIAS-met}& ESIAS-met/Melissa\\
    \hline 
    part. & cores & time (s) & core.s/part. & cores & time(s) & core.s/part. & resource usage ratio\\
 \hline
128 & 384 & 1062 & 3186 & 1536 & 267 & 3204 & 1.01\\
256 & 768 & 1062 & 3186 & 3072 & 317 & 3804 & 1.19\\
512 & 1536 & 1068 & 3204 & 6144 & 422 & 5064 & 1.58\\
1024 & 3072 & 1071 & 3213 & 12288 & 761 & 9132 & 2.84\\
    \hline
\end{tabular}
\caption{Comparing the resource usage (core.second/particle) per cycle  for Melissa and  ESIAS-met (file-based)  runs.}\label{tab:op3}
\end{table}

We compare Melissa with the  file-based approach ESIAS-met~\cite{lu_optimization_2022} using the same
simulation code \ac{wrf}~(V3.7.1) and the same  data set.
For the same number of particle, both approaches use a very different amount of cores (\Cref{tab:op3}).
With ESIAS-met each particle propagation requires to start a dedicated instance of WRF. Each time it includes  the cost from
loading and storing the particle state from/to a file.  At 1024 particles ESIAS-met uses $12 288$  cores while Melissa
just needs $3 072$ cores  as runners propagate several particles each. ESIAS-met  execution time  is thus shorter as highly parallelized, but
the resource usage (core.second/particle/cycle) is $2.84$ times  improved for Melissa due to  the combined strategies
developed to improve efficiency. The gain increases with the number of particles, showing that the  Melissa approach is particularly beneficial when targeting the large ensemble size.

\hide{
To assess the general performance of our approach, we tested it against
the state-of-the-art file-based (offline) approach ESIAS with
\ac{wrf}~(V3.7.1)~\cite{lu_optimization_2022}. ESIAS allows to set the
number of particles to include in the simulation, however, it does not
support particle virtualization (multiple particles assigned to one WRF
instance). Instead, the particles are executed in parallel with as many
instances of WRF as particles. An important result from this measurement
is, that the resources are used more efficiently in our approach. The
overall execution time is less for ESIAS, however, the core-hours per
cycle and particle are significantly higher (\Cref{tab:op3}). For
instance, the time average for a cycle with 1024 particles is about 1100
seconds using our implementation (result from the table multiplied by 4),
and 800 seconds for ESIAS. Hence, our approach takes 1.375 times longer
each cycle than ESIAS, however, ESIAS uses 4 times as many resources.
Besides the efficient utilization of resources, we observe an optimal
scaling for our implementation, in contrast to the poor scaling behavior
of ESIAS without particle virtualization. 
}

\hide{
\begin{table}[h!]
    \centering
   \scriptsize 
    \begin{tabular}{|c|c|c|c|c|c|}
    \hline
        Particles & Number Cores & CoreH/(cycle*particle) &
        time/(cycle*particle) & Speed-up \\
        ~ & Melissa/ESIAS & Melissa/ESIAS & Melissa/ESIAS & Melissa vs ESIAS \\ \hline
        128 & 384 /  1536 & 0.9962 / 0.8921 & 265.68  / 267.63 & 1.01 \\
        256 & 768 /  3072 & 0.9411 / 1.0591 & 265.72  / 317.74 & 1.20 \\
        512 & 1536 /  6144 & 0.918  / 1.407 & 267.06  / 422.13 & 1.58 \\
        1024 & 3072 / 12288 & 0.9067 / 2.5383 & 267.82  / 761.49 & 2.84 \\ \hline
    \end{tabular}
\caption{Option 3}\label{tab:op3}
\end{table} 
}

 \section{Discussion}\label{sec:discussion}

in~\Cref{sec:wrf_usecase} we derived that the total amount of data resulting
from 48 time-steps of particle filtering on the european domain with
\numprint{2555} particles accumulates to about 300~TiB.
Post processing this amount of data is challenging. Our framework could
be extended using in situ data processing techniques as
presented in Terraz et.\ al.~\cite{terraz-melissa-SC17}.

We only considered the case where the propagation time is longer than
the time for loading states from the \ac{pfs}; while applications where
propagations are shorter than loading the states would limit our
proposal efficiency, as we cannot further hide the I/O cost in that
case. On the other hand, we already have short propagation times in the
WRF context as we perform hourly resampling.  We chose this frequency
primarily to stress our proposal. Production runs usually do not require
such a high frequency, and rather have even longer propagation times as in
our experiments. However, to minimize transfer times further, we are
evaluating approaches leveraging node-local persistent storage as
globally shared storage layer. Solutions for this are readily available
in form of distributed ad~hoc file-systems~\cite{brinkmann_ad_2020} such
as BeeOND, GekkoFS, and BurstFS.  We have also experimented with
connecting the runners, establishing a peer to peer network, where
runners can exchange directly  the required states between each other. 

The particle propagation time in our experiments with \ac{wrf} is
relatively even, showing at most a $10\%$ variability.  Situations with
more variability are possible using different physics in WRF, with other
simulation codes, or, if runners execute on  heterogeneous resources,
some runners propagating faster than others by leveraging GPUs for
instance. Also use cases from other contexts such as Simulation Based
Inference (SIB) and ensemble classification, which can be performed
using our framework, might lead to vastly different propagation times.
Therefore, testing our framework under such conditions is an important
future work.

Our proposal currently relies on filters that do not compute internal
member state corrections.  Extending our approach to such  particle
filters~\cite{Leeuwen-particlefiltersurvey-2019} would possibly require
aggregating more than just the particle weights to the server. Exploring
the requirements to align our framework to such cases is the goal of a
future implementation of the particle filter that we propose. We
validated our proposal with the SIR particle filter, but many variations
exist and are active research topics~\cite{doucet2000sequential, surace2019avoid}.
One challenge is the exponentially
growing required particle number with the dimension of  the
problem~\cite{snyder_performance_2015,fearnhead_particle_2018}. This is
particularly acute for geoscience  use cases that, as in this paper,
work in high dimensional spaces. The
survey~\cite{Leeuwen-particlefiltersurvey-2019} gives an extensive
overlook of DA by particle filters  for geoscience and ways to cope with
dimensionality issues.

Particle filters, as  used here, require a
synchronization point at the end of each assimilation cycle. For our
framework, this is the major remaining source of inefficiency. Loosening
this requirement needs revisiting the particle filtering algorithm,
which constitutes an active    topic      of
research~\cite{Verge-parallelpartfilter-2015,Jasra-alivePartFilter-2013,Elvira-adaptingPartNumber-2016}.

 \section{Related Work}\label{sec:related-work}

The DA domain encompasses  a large variety of techniques and algorithms, like  nudging
\cite{Pauwels2001}, kriging \cite{Williams01Ab}, ensemble Kalman Filter \cite{Zhang2016}, ensemble
maximum likelihood filter \cite{Zhang2013}, or particle filter \cite{Leeuwen2003}. For an overview, we refer to~\cite{Asch-DABook-2016,Evensen-DABook-2009}. We focus here on statistical DA
relying on an  ensemble run of the model to compute a statistical estimator (co-variance matrix for EnKF,  PDF for particle filters).

To aggregate the data produced by all members (i.e., particles) two main groups of approaches are used. Either
the data is stored to files and then processed in a second step (off-line mode), or the data is processed on-line usually within a large MPI code in charge of running the members and data processing.
Frameworks relying on the off-line mode include  EnTK~\cite{Balasu-EnTK-IPDPS2018},
with the largest published DA use cases reaching \numprint{4096} members for a  molecular dynamics
application with an EnKF filter~\cite{balasubramanian-EnTk-2020}.  OpenDA also follows this model,
using NetCDF for data exchange with the NEMO code~\cite{van_velzen_openda-nemo_2016}. DART
supports both~\cite{anderson_data_2009}, with reports of large scale DA in off-line mode
in~\cite{Toye-FTschdDA-JCS2018} (about \numprint{1000}~members with an oceanic code),
or~\cite{Yashiro-1024DA-SC2020,Yashiro-2016}  (\numprint{1024} member, LETKF filter, \uu{6}{M}  Fugaku cores).
File based approaches have the benefit of their simplicity, providing  fault tolerance and
elasticity.  But  these solutions do not support member virtualization, state caching and
prefetching.  So starting or restarting a member requires to request a new resource allocation
launching  a new instance of the model code with  all the associated start-up costs.
Node-local persistent storage capabilities, for instance  with SSDs, can
store intermediate files, avoiding the PFS  to loosen the I/O bottleneck.
They are used for member state storage
in~\cite{Yashiro-1024DA-SC2020}, but without specific fault tolerance
mechanism.  So if a node fails and the node-local storage becomes
unavailable, the lost member states need to be recomputed.  Besides
leveraging the node storage for the distributed cache, using
node-storage rather than the parallel  file system as a globally shared
file system layer is one of our future goals.

The on-line mode  avoids the I/O bottleneck.
PDAF~\cite{Nerger-PDAF-2013}, which supports  both modes, has for
instance been used  on-line for the assimilation of observations into
the regional earth system model TerrSysMP. DA was based on  EnKF with up
to 256~members~\cite{kurtz_terrsysmppdaf_2016}.  ESIAS uses on-line DA
via particle filters with up to \numprint{4096} particles on a wind
power simulation on Europe~\cite{berndt_predictability_2018}.  Notice
that we work  with the same WRF component of ESIAS in this paper, using
a configuration on a similar domain but at higher spatial resolution and
with  more advanced and more time consuming physics. We also find ad hoc
MPI  codes for  on-line DA  as in ~\cite{miyoshi_10240-member_2014}
(atmospheric model, \numprint{10240} members, Local ENKF filter,
\numprint{4608} compute nodes). But all these MPI approaches  lead to
monolithic code without support for fault tolerance, elasticity or load
balancing. In \cite{bai2015particle}, the authors analyze various particle
propagation scheduling  but at  limited scale (6 compute nodes and 300 particles). 
We performed experiments on a similar architecture as our proposed one, but for \ac{enkf} instead of 
\ac{pf}~\cite{friedemann-melissaDA:2022}. We demonstrated fault-tolerance,
elasticity and scalability for experiments using up to \uu{16}{k} members, \uu{16}{k}
cores for DA with EnKF for the hydrology code Parflow. In contrast to
our novel proposal, \ac{enkf} requires a centralized filter update,
gathering the full ensemble of states at the central instance for the
assimilation of observations. In our novel proposal for \ac{pf}, we
exploit certain properties of particle filters to suppress the
server bottleneck and significantly reduce data movements.

 \section{Conclusion}\label{sec:conclusion}



In this article we proposed an architecture for handling very large ensembles for particle
filters. The architecture was designed to address the challenge of exascale computing that will
allow massive  ensemble runs~\cite{schulthess_reflecting_2019}.  The architecture is based on a server/runner model where runners
support a  distributed cache and virtualization of particle propagation, while the server aggregates the weights computed by the
runners and ensures the dynamic balancing of the work load.          Particle propagation is virtualized  so
the required number of runners is decoupled  from the particle number.  With the addition of a
distributed  checkpointing mechanism,  the architecture supports dynamic changes in the number of
runners during execution for fault tolerance and elasticity.  Experiments with the WRF weather
simulation code show that our framework can run at least \numprint{2555} particles on \numprint{20442} cores with a
$87\%$ scaling efficiency. Dynamic particle-propagation scheduling  and caching enable to avoid
$88\%$ of the global I/O operations.  Compared to the ESIAS file based approach, Melissa improves resource usage
$2.83$ times at $1024$ particles.


Future work  includes experimenting with  adaptive or localized particle  filters as well as
combining particle and Kalman filter.  We  also plan to extend the distributed cache  and
fault tolerance algorithm to fully avoid the centralized file system and only rely on node-local SSDs for particle storage.

\section*{Acknowledgement}

This project has received funding from the European Union's Horizon 2020 research and innovation program under grant agreement No 824158 (EoCoE-2).  This work was granted access to the HPC resources of IDRIS  under the allocation 2020-A8  A0080610366 attributed by GENCI (Grand Equipement National de Calcul Intensif). The authors gratefully acknowledge the Gauss Centre for Supercomputing e.V. (www.gauss-centre.eu) for funding this project by providing computing time through the John von Neumann Institute for Computing (NIC) on the GCS Supercomputer JUWELS at Jülich Supercomputing Centre (JSC). We acknowledge the access to the meteorological input data from the Meteocloud of SDL Climate Science, JSC. We also acknowledge PRACE for awarding us access to JUWELS at Jülich Supercomputing Centre (JSC), Germany.


\listoffixmes

\bibliographystyle{elsarticle-num}
\bibliography{references}

\end{document}